\documentclass{article}%
\usepackage{amssymb}
\usepackage{amsmath}
\usepackage{hyperref}
\usepackage{amsfonts}
\usepackage{graphicx}%
\providecommand{\U}[1]{\protect\rule{.1in}{.1in}}
\pdfoutput=1
\begin{document}

\title{Charged line segments and ellipsoidal equipotentials}
\author{T L Curtright$^{\S }$, N M Aden, X Chen, M J Haddad, S Karayev, D B Khadka,
and J Li\\Department of Physics, University of Miami\\Coral Gables, FL 33124-8046, USA\\$^{\S }${\small curtright@miami.edu}}
\maketitle

\begin{abstract}
This is a survey of the electrostatic potentials produced by charged
straight-line segments, in various numbers of spatial dimensions, with
comparisons between uniformly charged segments and those having non-uniform
linear charge distributions that give rise to ellipsoidal equipotentials
surrounding the segments. \ A uniform linear distribution of charge is
compatible with ellipsoidal equipotentials only for three dimensions. \ In
higher dimensions, the linear charge density giving rise to ellipsoidal
equipotentials is counter-intuitive --- the charge distribution has a maximum
at the center of the segment and vanishes at the ends of the segment. \ Only
in two dimensions is the continuous charge distribution intuitive --- for that
one case of ellipsoidal equipotentials, the charge is peaked at the ends of
the segment and minimized at the center.

\end{abstract}
\tableofcontents

\vfill

\section{Introduction}

It has become a widespread practice to study the physics of systems in various
numbers of spatial dimensions, not necessarily $D=3$. \ For instance, graphene
with $D=2$, and string or membrane theory with $D$ as high as $25$, are two
examples that immediately come to mind for both their experimental and
theoretical interest. \ Moreover, geometric ideas provide a common framework
used to pursue such studies. \ A pedagogical goal of this paper is to
encourage students to think along these lines in the context of a familiar
subject --- electrostatics.\newpage

For example, in three spatial dimensions a uniformly charged straight-line
segment gives rise to an electric potential $\Phi$ whose equipotential
surfaces are prolate ellipsoids of revolution about the segment, with the ends
of the segment providing the foci of the ellipsoid. \ As an immediate
consequence of the geometry for these prolate ellipsoidal equipotentials, the
associated electric field\ $\overrightarrow{E}=-\overrightarrow{\nabla}\Phi$
--- always normal to surfaces of constant $\Phi$ --- has at any observation
point a direction that bisects the angle formed by the pair of lines from the
observation point to each of the two foci of the ellipsoid. \ This beautiful
electrostatic example was presented by George Green in 1828 \cite{Green}, and
it has been discussed in many books since then \cite{Ferrers}-\cite{Zangwill}
including at least two texts from this century \cite{TSvK}. \ While the
straight-line segment is an idealization, nevertheless it provides insight
into the behavior of real thin-wire conductors, especially upon approximating
those real wires as very narrow, \textquotedblleft
needle-like\textquotedblright\ ellipsoids.

But as it turns out, the line segment problem in three dimensions is a very
special case, in some sense the most ideal of all possible electrostatic
worlds. \ In any \emph{other} dimension of space, uniformly charged segments
do \textbf{not} produce ellipsoidal equipotentials. \ Conversely, in any other
dimension, if the equipotentials are ellipsoidal about a linearly distributed
straight-line segment of charge, then that charge distribution can
\textbf{not} be uniform. \ 

One need look no farther than two dimensional systems to see clearly that
there are differences between uniformly charged segments and those with charge
distributed so as to produce ellipsoidal equipotentials. \ Indeed, $D=2$ is
the only ellipsoidal equipotential case which is intuitive in the sense that
the associated linear charge distribution has maxima at the ends of the
segment, as one might naively expect from the repulsive force between like
charges placed on a segment of a real, thin conductor at a finite potential.
\ In contrast, the distribution of charge needed to produce ellipsoidal
equipotentials is counter-intuitive in higher dimensions. \ To produce such
equipotentials for $D>3$ the charge distribution must have a maximum at the
center and vanish at the ends of the segment.

We begin our discussion in Section 2 in two dimensions, where the two types of
charged segments are readily analyzed. \ Then we compare and contrast
uniformly charged line segments with those admitting ellipsoidal
equipotentials for any number of spatial dimensions. \ As a preliminary, we
first discuss briefly in Section 3 the potential of a point charge in $D$
dimensions. \ We then use this information in Section 4 to compute the
potentials and electric fields for uniformly charged line segments. \ We
continue in Section 5 by considering systems with ellipsoidal equipotentials
in $D$\ dimensions. \ We then determine the linear charge distributions that
produce such potential configurations, and we find the remarkable result that
the linear charge density giving rise to ellipsoidal equipotentials is peaked
at the center of the segment, for $D>3$. \ 

The discussion of general $D$ affords the opportunity to illustrate how
continuous $D$ can be used as a mathematical device to regulate singular
behavior. \ This too is a widespread practice in theoretical physics. \ We use
$D$ in this way in Section 5 of the paper to interpolate continuously between
intuitive and counter-intuitive charge distributions for segments with
ellipsoidal equipotentials. \ 

Finally, in Section 6, we invoke well-known methods to indicate how the
various line segment results also give solutions to a class of electrostatic
boundary value problems where the charge is moved outward and distributed on
one of the equipotential surfaces that surrounds the original segment.

\section{Electrostatics in two dimensions}

The point-particle electric potential in 2D is well-known to be logarithmic.
\ For a point charge $Q$ located at the origin, up to a constant $R$ that sets
the distance scale,
\begin{equation}
\Phi_{\text{point}}\left(  \overrightarrow{r}\right)  =kQ\ln\left(
R/r\right)  \ ,\text{\ \ \ \ \ }\nabla^{2}\Phi_{\text{point}}\left(
\overrightarrow{r}\right)  =-2\pi kQ~\delta^{2}\left(  \overrightarrow{r}%
\right)  \ , \label{2DPointCharge}%
\end{equation}
where $k$ is the two dimensional analogue of Coulomb's constant. \ 

By linear superposition, a finite length, uniformly charged line segment, with
constant charge/length $\lambda$ distributed on the $x$ axis for $-L/2\leq
x\leq L/2$, produces a potential
\begin{equation}
\Phi_{\text{line}}\left(  x,y\right)  =-k\lambda\left.  \left(  u\ln
\sqrt{u^{2}+y^{2}}-u+y\arctan\frac{u}{y}\right)  \right\vert _{u=-x-L/2}%
^{u=-x+L/2}+k\lambda L\ln\left(  R\right)  \ .
\end{equation}
This result may be established by integrating the contributions of
infinitesimal point-like bits of charge that make up the segment, using
(\ref{2DPointCharge}) and the indefinite integral
\begin{equation}
\int\ln\left(  \sqrt{u^{2}+y^{2}}\right)  du=u\ln\left(  \sqrt{u^{2}+y^{2}%
}\right)  -u+y\arctan\frac{u}{y}\ .
\end{equation}
Written out in full,%
\begin{align}
\Phi_{\text{line}}\left(  x,y\right)   &  =k\lambda\left(  L+y\arctan\left(
\frac{x-\frac{1}{2}L}{y}\right)  -y\arctan\left(  \frac{x+\frac{1}{2}L}%
{y}\right)  \right) \label{UniformLine2D}\\
&  +\left(  x-\frac{1}{2}L\right)  k\lambda\ln\left(  \frac{1}{R}\sqrt{\left(
x-\frac{1}{2}L\right)  ^{2}+y^{2}}\right)  -\left(  x+\frac{1}{2}L\right)
k\lambda\ln\left(  \frac{1}{R}\sqrt{\left(  x+\frac{1}{2}L\right)  ^{2}+y^{2}%
}\right)  \ .\nonumber
\end{align}
A plot of the potential surface shows the essential features.%
%TCIMACRO{\FRAME{dtbpFU}{6.7428in}{4.4855in}{0pt}{\Qcb{Figure 1: \ Potential
%surface for a uniformly charged line segment in 2D.}}{}%
%{eurojphyslinesegmentsv2__1.pdf}{\special{ language "Scientific Word";
%type "GRAPHIC";  maintain-aspect-ratio TRUE;  display "USEDEF";
%valid_file "F";  width 6.7428in;  height 4.4855in;  depth 0pt;
%original-width 6.4118in;  original-height 4.2557in;  cropleft "0";
%croptop "1";  cropright "1";  cropbottom "0";
%filename 'EuroJPhysLineSegmentsV2__1.pdf';file-properties "XNPEU";}} }%
%BeginExpansion
\begin{center}
\includegraphics[
height=4.4855in,
width=6.7428in
]%
{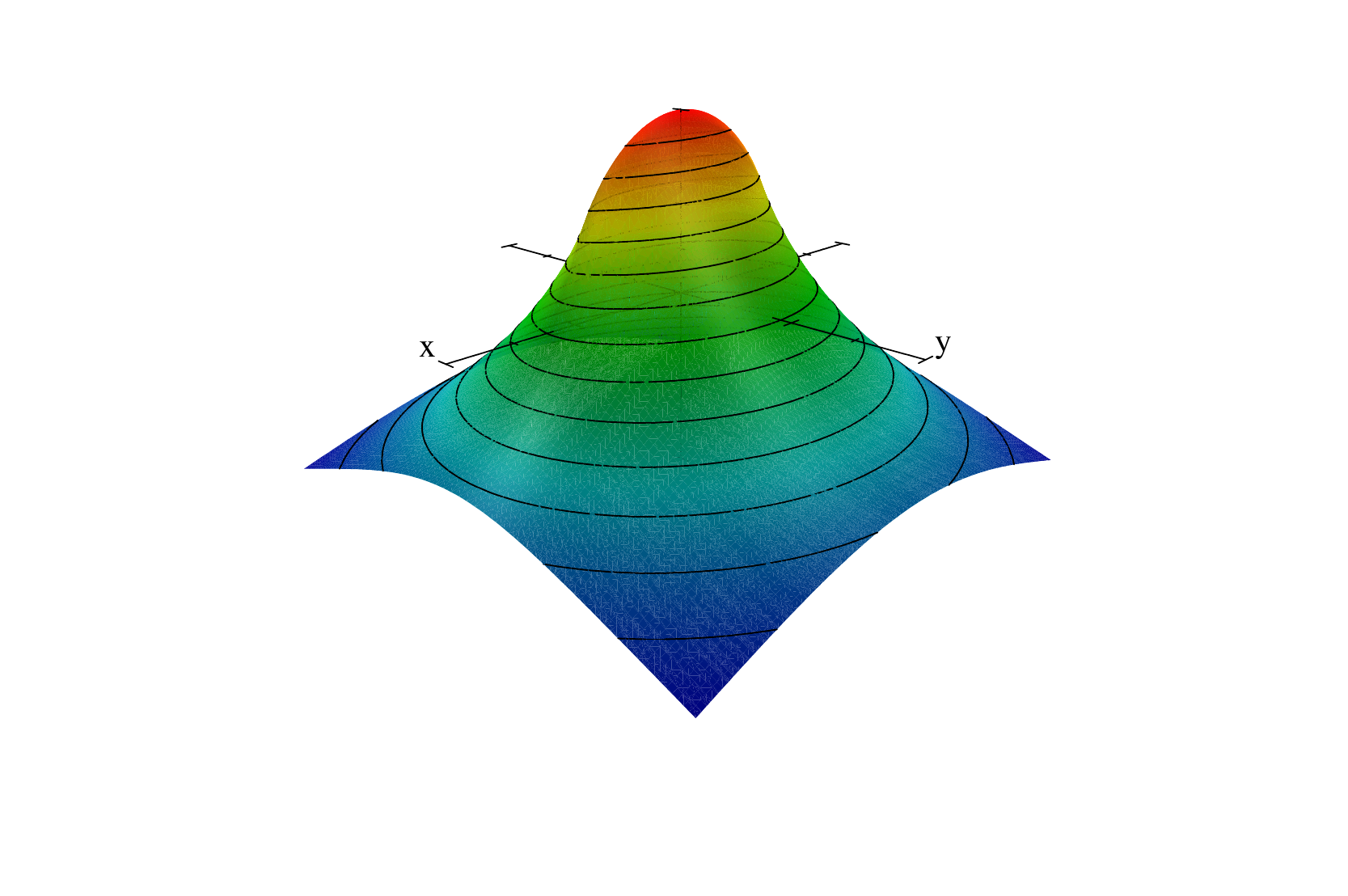}%
\\
Figure 1: \ Potential surface for a uniformly charged line segment in 2D.
\end{center}
%EndExpansion

The top of the potential surface is curved and not a straight line, indicating
that the charged segment itself is \emph{not} an equipotential. \ This follows
analytically from (\ref{UniformLine2D}). \ Although points on the segment are
not at the \emph{same} potential, they are all at \emph{finite} values of the
potential, for $D=2$. \ Explicitly, for $y=0$ and all $x$,%
\begin{equation}
\Phi_{\text{line}}\left(  x,0\right)  =k\lambda L-k\lambda\left(  \tfrac{1}%
{2}L+x\right)  \ln\left(  \left\vert \tfrac{1}{2}L+x\right\vert /R\right)
-k\lambda\left(  \tfrac{1}{2}L-x\right)  \ln\left(  \left\vert \tfrac{1}%
{2}L-x\right\vert /R\right)  \ .
\end{equation}
We plot $f\left(  x\right)  =\Phi_{\text{line}}\left(  x,0\right)  /k\lambda$
versus $x$ to show the shape of the potential along the $x$-axis, for $L=2$
and $R=1$.%
%TCIMACRO{\FRAME{dtbpFU}{4.1458in}{2.7576in}{0pt}{\Qcb{Figure 2: \ Uniformly
%charged line segment potential for $D=2$, along the $x$-axis.}}{}%
%{eurojphyslinesegmentsv2__2.pdf}{\special{ language "Scientific Word";
%type "GRAPHIC";  maintain-aspect-ratio TRUE;  display "USEDEF";
%valid_file "F";  width 4.1458in;  height 2.7576in;  depth 0pt;
%original-width 3.9319in;  original-height 2.606in;  cropleft "0";
%croptop "1";  cropright "1";  cropbottom "0";
%filename 'EuroJPhysLineSegmentsV2__2.pdf';file-properties "XNPEU";}} }%
%BeginExpansion
\begin{center}
\includegraphics[
height=2.7576in,
width=4.1458in
]%
{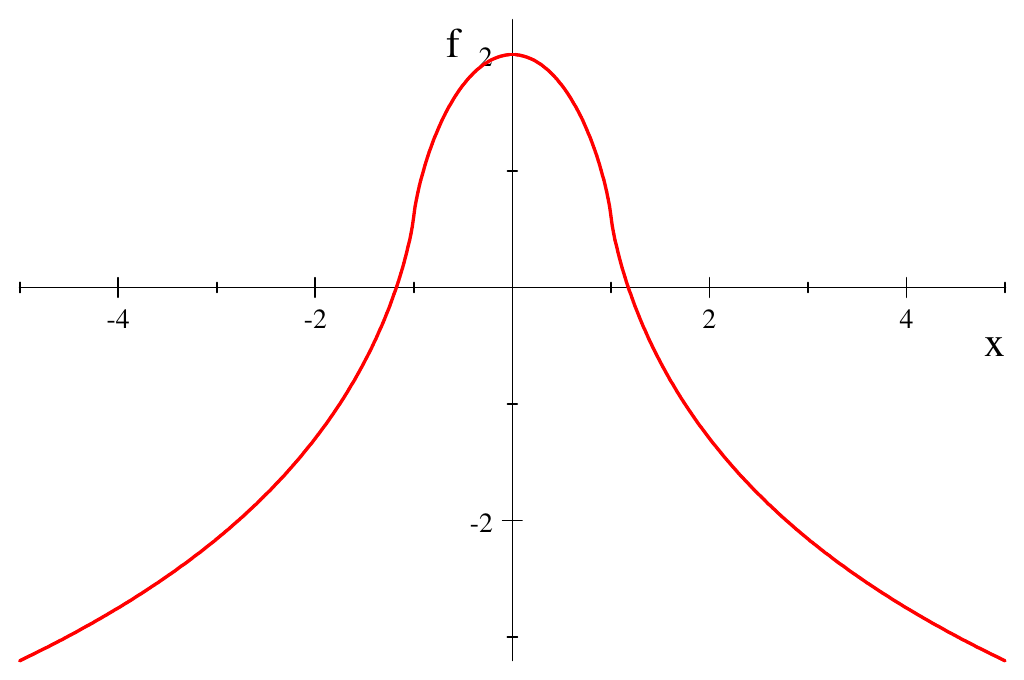}%
\\
Figure 2: \ Uniformly charged line segment potential for $D=2$, along the
$x$-axis.
\end{center}
%EndExpansion

On the other hand, transverse to the $x$-axis the potential has a
discontinuous slope across the line segment. \ That is to say, the electric
field normal to the line segment, $E_{y}$, is discontinuous due to the
presence of the charge density on the segment. \ For example, for $x=0$ this
transverse profile is given by%
\begin{equation}
\Phi\left(  0,y\right)  =k\lambda\left[  L-L\ln\left(  \sqrt{\tfrac{1}{4}%
L^{2}+y^{2}}/R\right)  -2y\arctan\left(  \tfrac{1}{2}L/y\right)  \right]  \ .
\end{equation}
We also plot $g\left(  y\right)  =\Phi_{\text{line}}\left(  0,y\right)
/k\lambda$ versus $y$ to show the shape of the potential along the $y$-axis,
for $L=2$ and $R=1$.%
%TCIMACRO{\FRAME{dtbpFU}{4.1458in}{2.7576in}{0pt}{\Qcb{Figure 3: \ Uniformly
%charged line segment potential for $D=2$, along the $y$-axis.}}{}%
%{eurojphyslinesegmentsv2__3.pdf}{\special{ language "Scientific Word";
%type "GRAPHIC";  maintain-aspect-ratio TRUE;  display "USEDEF";
%valid_file "F";  width 4.1458in;  height 2.7576in;  depth 0pt;
%original-width 3.9319in;  original-height 2.606in;  cropleft "0";
%croptop "1";  cropright "1";  cropbottom "0";
%filename 'EuroJPhysLineSegmentsV2__3.pdf';file-properties "XNPEU";}} }%
%BeginExpansion
\begin{center}
\includegraphics[
height=2.7576in,
width=4.1458in
]%
{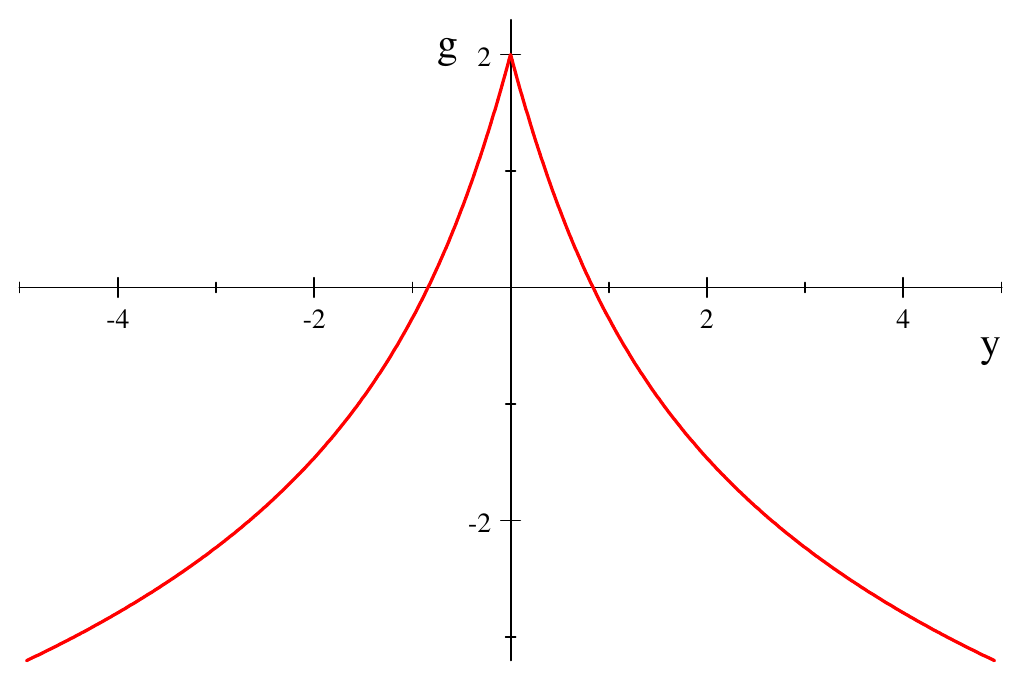}%
\\
Figure 3: \ Uniformly charged line segment potential for $D=2$, along the
$y$-axis.
\end{center}
%EndExpansion

Moreover, the equipotentials are \emph{not} ellipses surrounding the segment
on the $xy$-plane, as is especially clear for points close to the segment.
\ The potential contours actually \emph{intersect} the segment. \ A view of
the potential contours from \emph{below} the potential surface shows these
features graphically. \ (See the plot to follow. \ But note the view in that
plot is actually an orthogonal projection of the contours onto the $xy$-plane,
and not the true perspective of an observer on the potential axis a finite
distance below that plane.)%
%TCIMACRO{\FRAME{dtbpFU}{6.7428in}{4.4855in}{0pt}{\Qcb{Figure 4: \ Potential
%contours for a uniformly charged line segment in 2D.}}{}%
%{eurojphyslinesegmentsv2__4.pdf}{\special{ language "Scientific Word";
%type "GRAPHIC";  maintain-aspect-ratio TRUE;  display "USEDEF";
%valid_file "F";  width 6.7428in;  height 4.4855in;  depth 0pt;
%original-width 6.4118in;  original-height 4.2557in;  cropleft "0";
%croptop "1";  cropright "1";  cropbottom "0";
%filename 'EuroJPhysLineSegmentsV2__4.pdf';file-properties "XNPEU";}} }%
%BeginExpansion
\begin{center}
\includegraphics[
height=4.4855in,
width=6.7428in
]%
{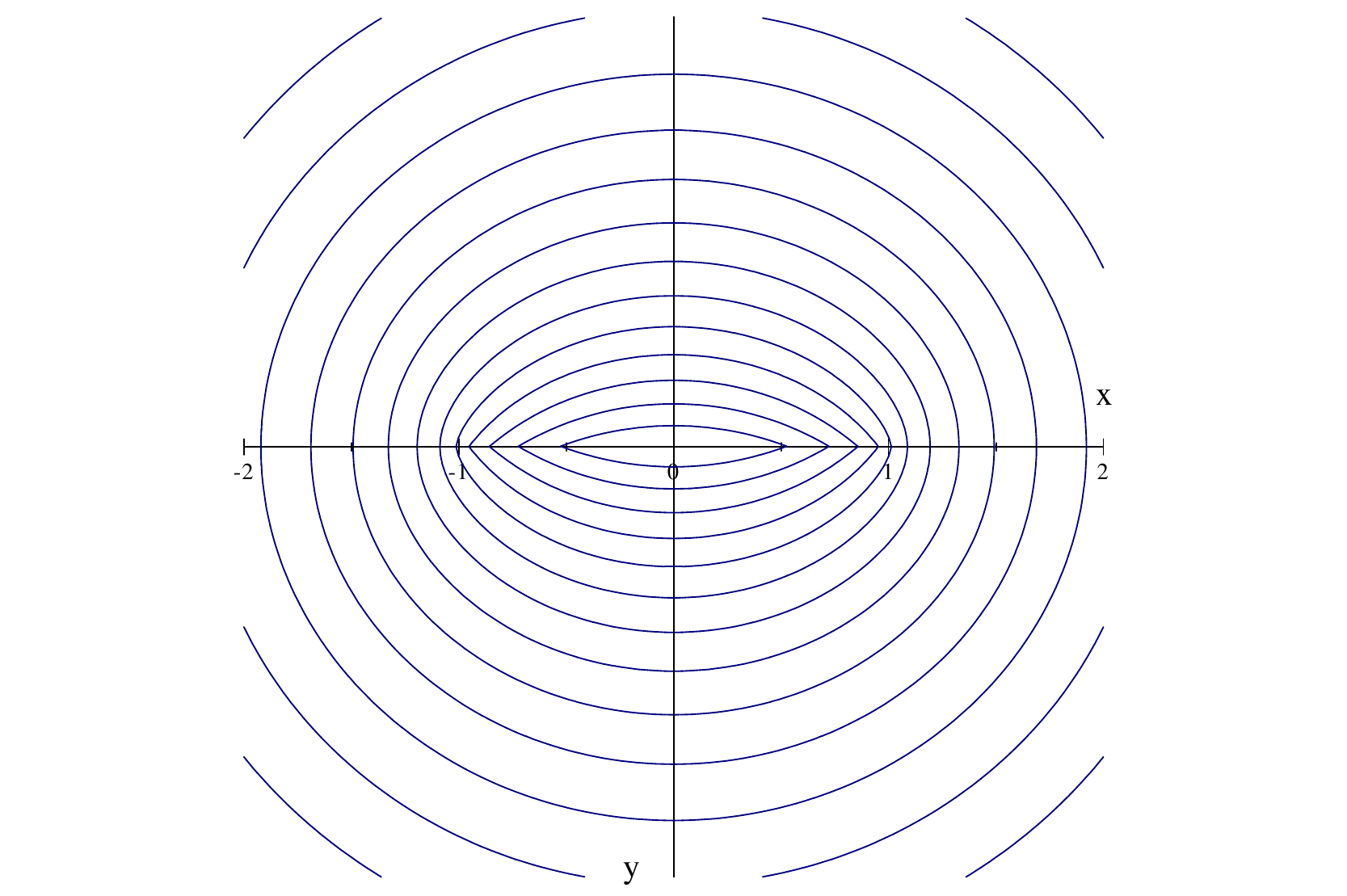}%
\\
Figure 4: \ Potential contours for a uniformly charged line segment in 2D.
\end{center}
%EndExpansion

The components of the electric field produced by the segment are given by%
\begin{align}
E_{x}\left(  x,y\right)   &  =-\frac{\partial}{\partial x}\Phi_{\text{line}%
}\left(  x,y\right)  =k\lambda\ln\left(  \sqrt{\frac{\left(  x+\frac{1}%
{2}L\right)  ^{2}+y^{2}}{\left(  x-\frac{1}{2}L\right)  ^{2}+y^{2}}}\right)
\ ,\\
E_{y}\left(  x,y\right)   &  =-\frac{\partial}{\partial y}\Phi_{\text{line}%
}\left(  x,y\right)  =k\lambda\arctan\left(  \frac{L+2x}{2y}\right)
+k\lambda\arctan\left(  \frac{L-2x}{2y}\right)  \ ,
\end{align}
where we have used $\frac{d}{du}\arctan u=\frac{1}{u^{2}+1}$. \ Note that
$\overrightarrow{E}\left(  x,y\right)  $ is independent of the scale $R$,
since changing $R$ just amounts to adding a constant to the potential. \ But
of course $\overrightarrow{E}\left(  x,y\right)  $ does depend on $L$ as this
sets the physical length scale for the system.

As a check, in the units we have chosen the charge density along the line
follows from $\int\overrightarrow{\nabla}\cdot\overrightarrow{E}~dxdy=2\pi
k\int\rho~dxdy$, integrated over a horizontal rectangle containing an
infinitesimal portion of the $x$-axis, as the height of the rectangle is taken
to zero. \ Thus, using $\lim\limits_{z\rightarrow\pm\infty}\arctan z=\pm\pi
/2$,%
\begin{equation}
2\pi k~\lambda\left(  x\right)  =2\lim_{y\rightarrow0}E_{y}\left(  x,y\right)
=2k\lambda\times\left\{
\begin{array}
[c]{cl}%
0 & \text{ \ \ if \ \ }x>L/2\\
\pi & \text{ \ \ if \ \ }-L/2<x<L/2\\
0 & \text{ \ \ if \ \ }x<-L/2
\end{array}
\right.  \ . \label{UniformCharge2DDerivation}%
\end{equation}
That is to say, the charge/length between $-L/2$ and $L/2$ is the constant
$\lambda$, as expected. \ Elsewhere, the charge density vanishes, as follows
from $\nabla^{2}\Phi_{\text{line}}\left(  \overrightarrow{r}\right)  =0$ for
all points not coincident with the segment.

It is instructive to make a vector field plot of $\left(  E_{x}\left(
x,y\right)  ,E_{y}\left(  x,y\right)  \right)  $, especially near the charged
segment. \ Again it is evident graphically that the segment itself is not an
equipotential, since the electric field lines are not perpendicular to the
segment as the $x$-axis is approached for $-L/2<x<L/2$, except at the single
point $x=0$. \ Here are vector field plots for $L=2$.%
%TCIMACRO{\FRAME{dtbpF}{6.2665in}{6.2648in}{0pt}{}{}%
%{eurojphyslinesegmentsv2__5.pdf}{\special{ language "Scientific Word";
%type "GRAPHIC";  display "USEDEF";  valid_file "F";  width 6.2665in;
%height 6.2648in;  depth 0pt;  original-width 5.9983in;
%original-height 5.9817in;  cropleft "0";  croptop "1";  cropright "1";
%cropbottom "0";
%filename 'EuroJPhysLineSegmentsV2__5.pdf';file-properties "XNPEU";}} }%
%BeginExpansion
\begin{center}
\includegraphics[
height=6.2648in,
width=6.2665in
]%
{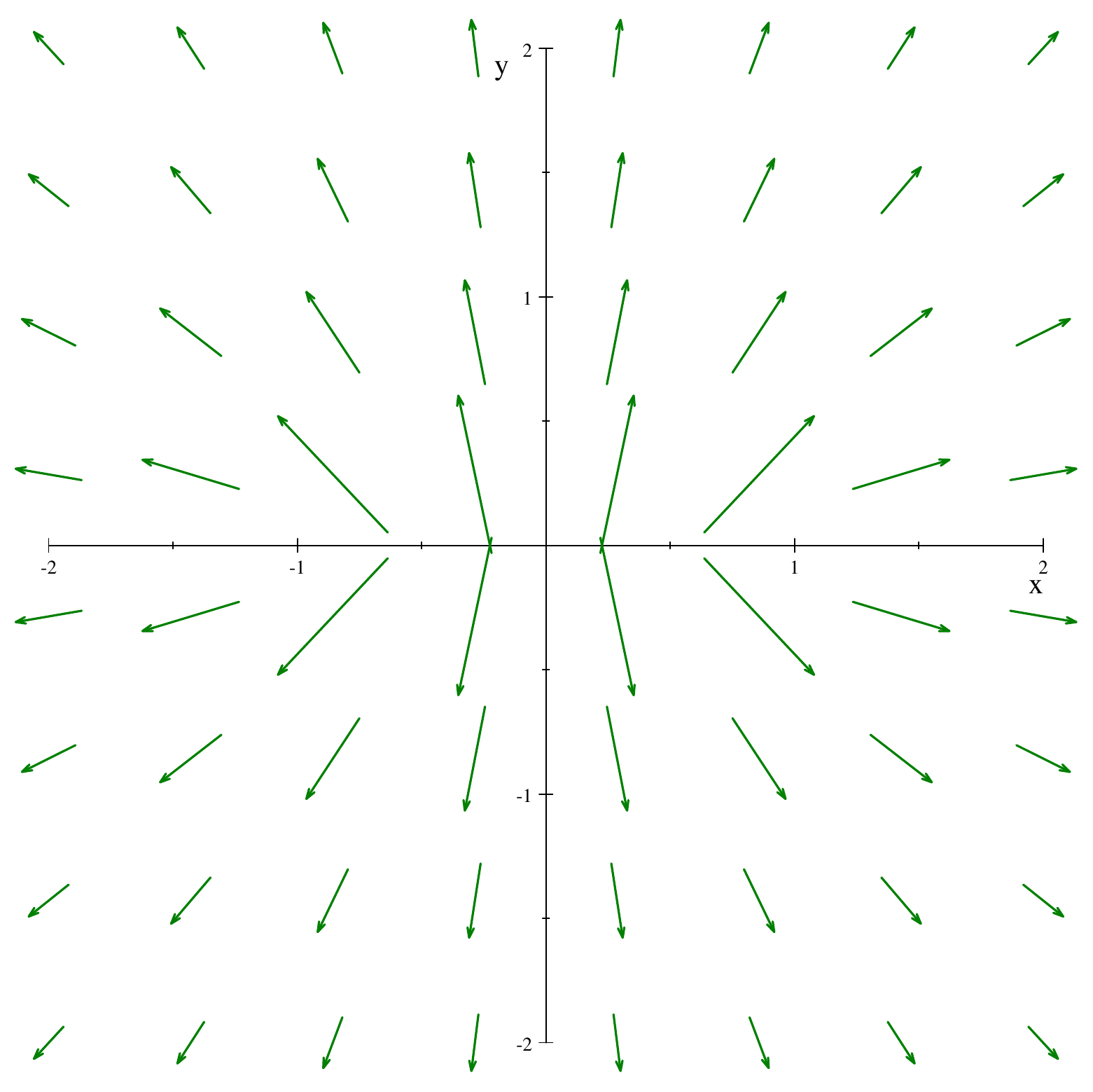}%
\end{center}
%EndExpansion%
%TCIMACRO{\FRAME{dtbpFU}{6.7428in}{2.064in}{0pt}{\Qcb{Figures 5 \& 6: \ Vector
%field plots of $\protect\overrightarrow{E}$ for a uniformly charged line
%segment situated between $x=-1$ and $x=+1$. \ The field is evaluated at the
%center of each arrow.}}{}{eurojphyslinesegmentsv2__6.pdf}%
%{\special{ language "Scientific Word";  type "GRAPHIC";
%maintain-aspect-ratio TRUE;  display "USEDEF";  valid_file "F";
%width 6.7428in;  height 2.064in;  depth 0pt;  original-width 6.4118in;
%original-height 1.9444in;  cropleft "0";  croptop "1";  cropright "1";
%cropbottom "0";
%filename 'EuroJPhysLineSegmentsV2__6.pdf';file-properties "XNPEU";}} }%
%BeginExpansion
\begin{center}
\includegraphics[
height=2.064in,
width=6.7428in
]%
{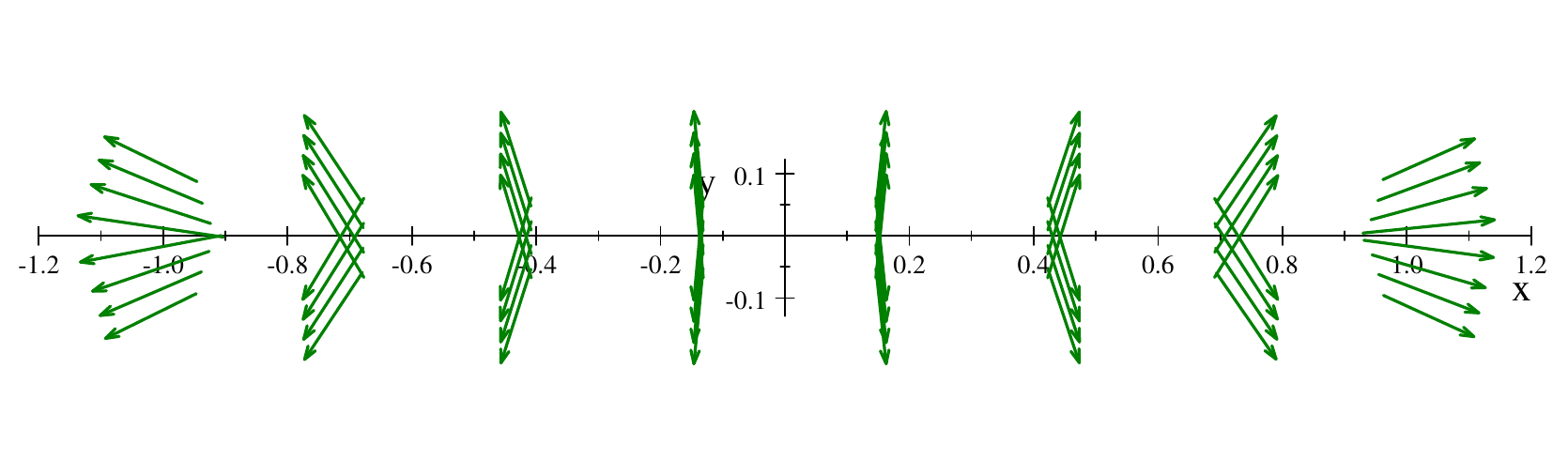}%
\\
Figures 5 \& 6: \ Vector field plots of $\protect\overrightarrow{E}$ for a
uniformly charged line segment situated between $x=-1$ and $x=+1$. \ The field
is evaluated at the center of each arrow.
\end{center}
%EndExpansion

In contrast to the uniformly charged segment in 2D, consider a distribution of
charge along the segment such that equipotentials \emph{are} ellipsoidal.
\ The relevant charge distribution turns out to be%
\begin{equation}
\lambda\left(  x\right)  =\frac{2\lambda L}{\pi}\frac{1}{\sqrt{L^{2}-4x^{2}}%
}\left\{
\begin{array}
[c]{cl}%
0 & \text{ \ \ if \ \ }x>L/2\\
1 & \text{ \ \ if \ \ }-L/2<x<L/2\\
0 & \text{ \ \ if \ \ }x<-L/2
\end{array}
\right.  \ , \label{Lambda(x)2D}%
\end{equation}
as we shall confirm in the following. \ Since $\int_{-1}^{1}\frac{1}%
{\sqrt{1-s^{2}}}\,ds=\pi$, the total charge on the segment is still
$Q=\int_{-L/2}^{L/2}\lambda\left(  x\right)  \,dx=\lambda L$, the same as for
the uniformly charged case. \ 

The corresponding potential is now \
\begin{equation}
\Phi_{\text{line}}\left(  x,y\right)  =k\lambda L\ln\left(  \frac{4R}%
{s+\sqrt{s^{2}-L^{2}}}\right)  \ , \label{EllipsoidalLine2D}%
\end{equation}
where $s$ is a sum of two distances, from the observation point $\left(
x,y\right)  $ to each of the two ends of the segment. \ That is,%
\begin{equation}
s=r_{-}+r_{+}\ ,\ \ \ \ \ r_{\pm}=\sqrt{\left(  x\pm\frac{1}{2}~L\right)
^{2}+y^{2}}\ .
\end{equation}
Because the positional dependence of the potential is given entirely by $s$,
the equipotentials are ellipses on the $xy$-plane, with the ends of the
segment serving as the foci of each equipotential ellipse. \ Again, we plot
the potential surface.%
%TCIMACRO{\FRAME{dtbpFU}{6.749in}{4.4899in}{0pt}{\Qcb{Figure 7: \ Potential
%surface for a non-uniformly charged line segment in 2D, with $\lambda\left(
%x\right)  =\frac{2\lambda L/\pi}{\sqrt{L^{2}-4x^{2}}}$.}}{}%
%{eurojphyslinesegmentsv2__7.pdf}{\special{ language "Scientific Word";
%type "GRAPHIC";  maintain-aspect-ratio TRUE;  display "USEDEF";
%valid_file "F";  width 6.749in;  height 4.4899in;  depth 0pt;
%original-width 6.8502in;  original-height 4.5467in;  cropleft "0";
%croptop "1";  cropright "1";  cropbottom "0";
%filename 'EuroJPhysLineSegmentsV2__7.pdf';file-properties "XNPEU";}} }%
%BeginExpansion
\begin{center}
\includegraphics[
height=4.4899in,
width=6.749in
]%
{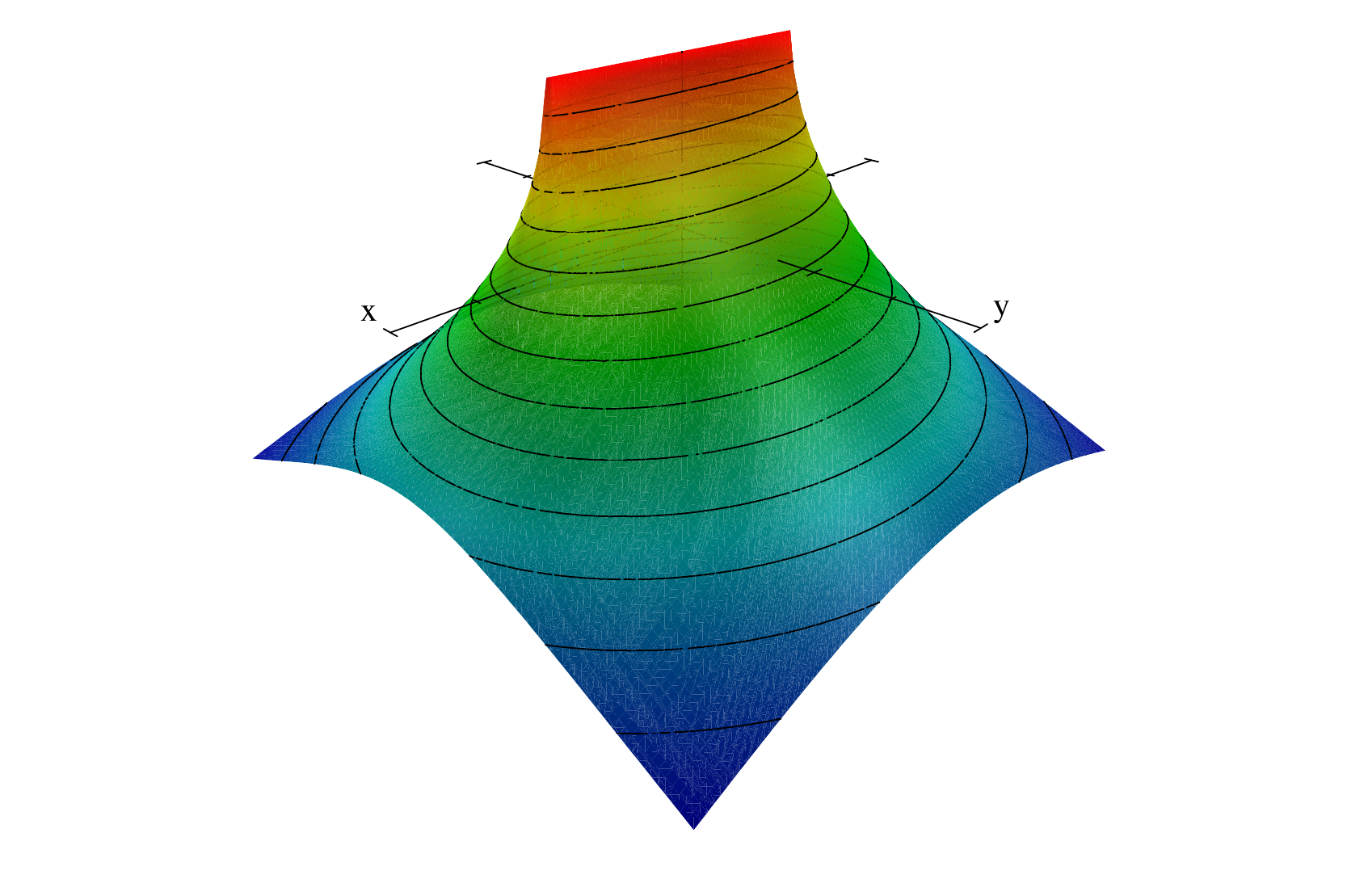}%
\\
Figure 7: \ Potential surface for a non-uniformly charged line segment in 2D,
with $\lambda\left(  x\right)  =\frac{2\lambda L/\pi}{\sqrt{L^{2}-4x^{2}}}$.
\end{center}
%EndExpansion

The top of the potential surface is now a straight line, indicating that the
charged segment itself \emph{is} an equipotential, and the points on the
segment are at a \emph{finite} potential for $D=2$, namely, $\Phi=k\lambda
L\ln\left(  4R/L\right)  $. \ This follows analytically from
(\ref{EllipsoidalLine2D}). \ Explicitly, for $y=0$ and all $x$,%
\begin{equation}
\Phi_{\text{line}}\left(  x,0\right)  =k\lambda L\ln\left(  \frac
{4R}{\left\vert x-\frac{1}{2}~L\right\vert +\left\vert x+\frac{1}%
{2}~L\right\vert +\sqrt{2\left(  x^{2}-\frac{1}{4}L^{2}+\left\vert x^{2}%
-\frac{1}{4}~L^{2}\right\vert \right)  }}\right)  \ .
\end{equation}
We plot $f\left(  x\right)  =\frac{1}{k\lambda}~\Phi_{\text{line}}\left(
x,0\right)  $ versus $x$ to show the shape of the potential along the
$x$-axis, for $L=2$ and $R=1$.%
%TCIMACRO{\FRAME{dtbpFU}{4.1493in}{2.7603in}{0pt}{\Qcb{Figure 8:
%\ Non-uniformly charged line segment potential for $D=2$, along the $x$%
%-axis.}}{}{eurojphyslinesegmentsv2__8.pdf}%
%{\special{ language "Scientific Word";  type "GRAPHIC";
%maintain-aspect-ratio TRUE;  display "USEDEF";  valid_file "F";
%width 4.1493in;  height 2.7603in;  depth 0pt;  original-width 4.2007in;
%original-height 2.7842in;  cropleft "0";  croptop "1";  cropright "1";
%cropbottom "0";
%filename 'EuroJPhysLineSegmentsV2__8.pdf';file-properties "XNPEU";}} }%
%BeginExpansion
\begin{center}
\includegraphics[
height=2.7603in,
width=4.1493in
]%
{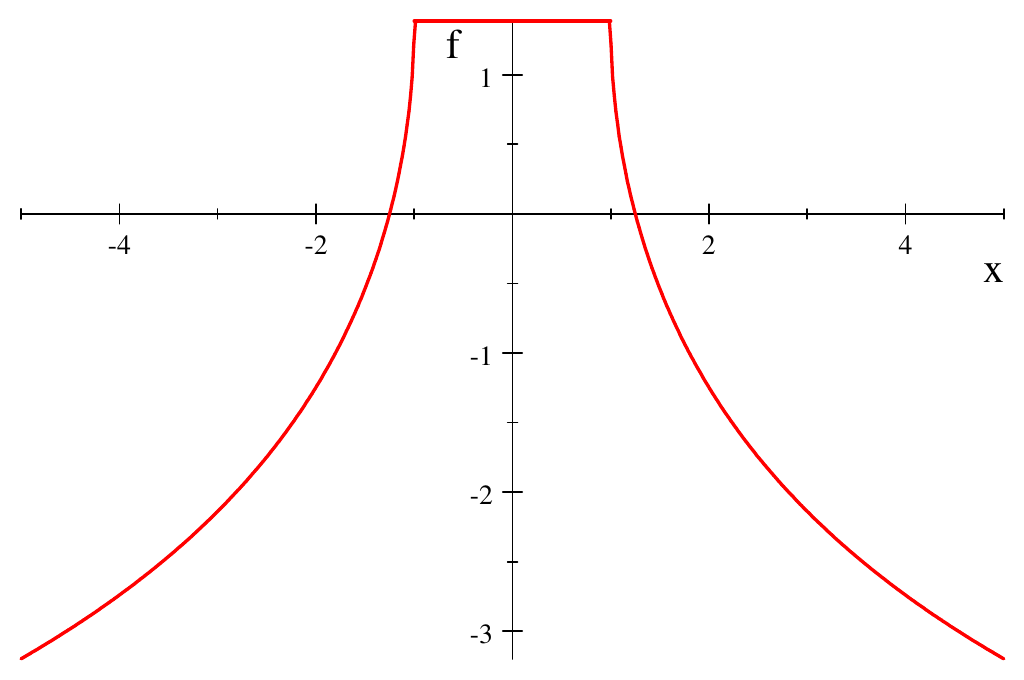}%
\\
Figure 8: \ Non-uniformly charged line segment potential for $D=2$, along the
$x$-axis.
\end{center}
%EndExpansion
Transverse to the $x$-axis, the potential again has a discontinuous slope
across the line segment due to the presence of the charge density on the
segment. \ For example, for $x=0$ this transverse profile is given by%
\begin{equation}
\Phi_{\text{line}}\left(  0,y\right)  =Lk\lambda\ln\left(  \frac{4R}%
{\sqrt{L^{2}+4y^{2}}+2\left\vert y\right\vert }\right)  \ .
\end{equation}
We plot $g\left(  y\right)  =\frac{1}{k\lambda}~\Phi_{\text{line}}\left(
0,y\right)  $ versus $y$ to show the shape of the potential along the
$y$-axis, for $L=2$ and $R=1$.%
%TCIMACRO{\FRAME{dtbpFU}{4.1493in}{2.7603in}{0pt}{\Qcb{Figure 9:
%\ Non-uniformly charged line segment potential for $D=2$, along the $y$%
%-axis.}}{}{eurojphyslinesegmentsv2__9.pdf}%
%{\special{ language "Scientific Word";  type "GRAPHIC";
%maintain-aspect-ratio TRUE;  display "USEDEF";  valid_file "F";
%width 4.1493in;  height 2.7603in;  depth 0pt;  original-width 4.2007in;
%original-height 2.7842in;  cropleft "0";  croptop "1";  cropright "1";
%cropbottom "0";
%filename 'EuroJPhysLineSegmentsV2__9.pdf';file-properties "XNPEU";}} }%
%BeginExpansion
\begin{center}
\includegraphics[
height=2.7603in,
width=4.1493in
]%
{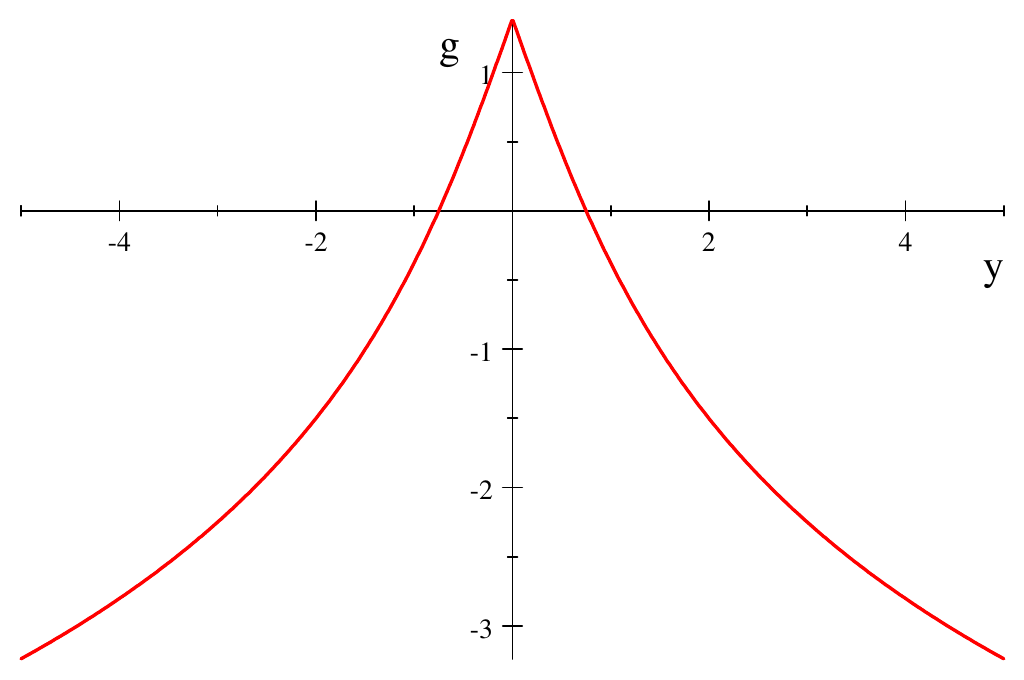}%
\\
Figure 9: \ Non-uniformly charged line segment potential for $D=2$, along the
$y$-axis.
\end{center}
%EndExpansion

A view of the potential contours from below the potential surface shows the
equipotential ellipses. \ (But again note the view in that plot is actually an
orthogonal projection of the contours onto the $xy$-plane, and not the true
perspective of an observer on the potential axis a finite distance below that
plane.)%
%TCIMACRO{\FRAME{dtbpFU}{6.749in}{4.4899in}{0pt}{\Qcb{Figure 10: \ Ellipsoidal
%potential contours for a non-uniformly charged line segment in 2D.}}%
%{}{eurojphyslinesegmentsv2__10.pdf}{\special{ language "Scientific Word";
%type "GRAPHIC";  maintain-aspect-ratio TRUE;  display "USEDEF";
%valid_file "F";  width 6.749in;  height 4.4899in;  depth 0pt;
%original-width 6.8502in;  original-height 4.5467in;  cropleft "0";
%croptop "1";  cropright "1";  cropbottom "0";
%filename 'EuroJPhysLineSegmentsV2__10.pdf';file-properties "XNPEU";}} }%
%BeginExpansion
\begin{center}
\includegraphics[
height=4.4899in,
width=6.749in
]%
{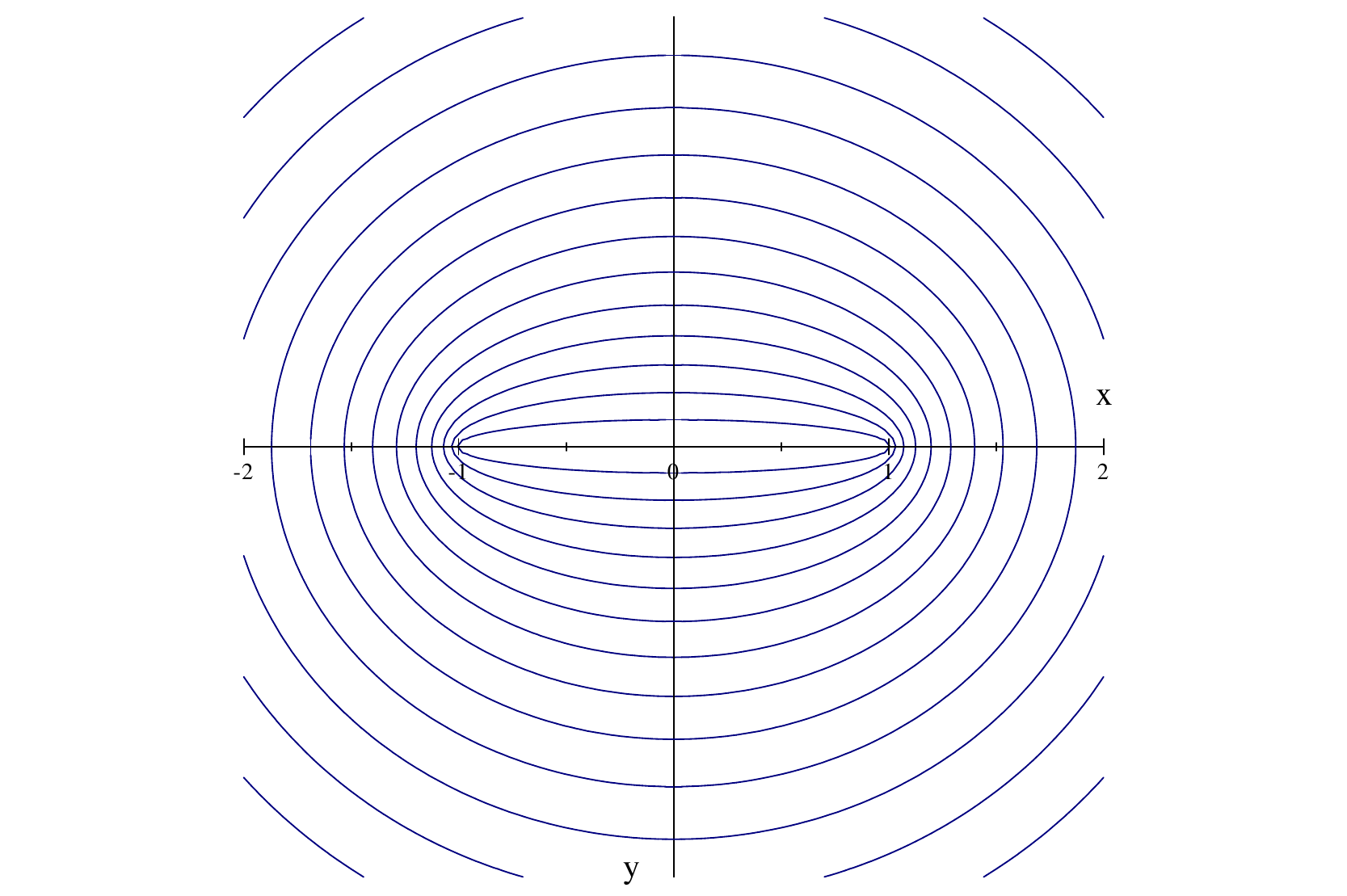}%
\\
Figure 10: \ Ellipsoidal potential contours for a non-uniformly charged line
segment in 2D.
\end{center}
%EndExpansion
\ 

The components of the electric field are now%
\begin{align}
E_{x}\left(  x,y\right)   &  =-\frac{\partial}{\partial x}\Phi_{\text{line}%
}\left(  x,y\right)  =\frac{k\lambda L}{\sqrt{s^{2}-L^{2}}}\frac{\partial
s}{\partial x}\ ,\\
E_{y}\left(  x,y\right)   &  =-\frac{\partial}{\partial y}\Phi_{\text{line}%
}\left(  x,y\right)  =\frac{k\lambda L}{\sqrt{s^{2}-L^{2}}}\frac{\partial
s}{\partial y}\ ,
\end{align}
whose final form in rectangular coordinates follows from
\begin{equation}
\frac{\partial s}{\partial x}=\frac{x-\frac{1}{2}~L}{r_{-}}+\frac{x+\frac
{1}{2}~L}{r_{+}}=\frac{1}{r_{+}r_{-}}\left(  sx-\frac{1}{2}~\left(
r_{+}-r_{-}\right)  L\right)  \ ,\ \ \ \frac{\partial s}{\partial y}=\frac
{y}{r_{-}}+\frac{y}{r_{+}}=\frac{1}{r_{+}r_{-}}~sy\ .
\end{equation}
In terms of $x,\ y,\ r_{+},$ and $r_{-}$,%
\begin{align}
E_{x}\left(  x,y\right)   &  =\frac{k\lambda L}{r_{+}r_{-}\sqrt{\left(
r_{-}+r_{+}\right)  ^{2}-L^{2}}}\left(  \left(  x-\frac{1}{2}~L\right)
r_{+}+\left(  x+\frac{1}{2}~L\right)  r_{-}\right)  \ ,\\
E_{y}\left(  x,y\right)   &  =\frac{k\lambda L}{r_{+}r_{-}\sqrt{\left(
r_{-}+r_{+}\right)  ^{2}-L^{2}}}\left(  r_{-}+r_{+}\right)  y\ .
\end{align}
Writing everything out in terms of $x$ and $y$ gives%
\begin{equation}
r_{+}r_{-}=\sqrt{\left(  \frac{1}{4}~L^{2}+x^{2}+y^{2}\right)  ^{2}-L^{2}%
x^{2}}\ ,\ \ \ \left(  r_{-}+r_{+}\right)  ^{2}-L^{2}=2\left(  x^{2}%
+y^{2}-\frac{1}{4}L^{2}+\sqrt{\left(  \frac{1}{4}~L^{2}+x^{2}+y^{2}\right)
^{2}-L^{2}x^{2}}\right)  \ .
\end{equation}
The components of $\overrightarrow{E}$\ are not elegant in these coordinates
\cite{Footnote1} but their properties are fully encoded and amenable to
machine computation.%
\begin{align}
E_{x}\left(  x,y\right)   &  =\frac{k\lambda L}{\sqrt{2}}\frac{\left(
x+\frac{1}{2}~L\right)  \sqrt{\left(  x-\frac{1}{2}~L\right)  ^{2}+y^{2}%
}+\left(  x-\frac{1}{2}~L\right)  \sqrt{\left(  x+\frac{1}{2}~L\right)
^{2}+y^{2}}}{\sqrt{\left(  \frac{1}{4}~L^{2}+x^{2}+y^{2}\right)  ^{2}%
-L^{2}x^{2}}\sqrt{x^{2}+y^{2}-\frac{1}{4}L^{2}+\sqrt{\left(  \frac{1}{4}%
~L^{2}+x^{2}+y^{2}\right)  ^{2}-L^{2}x^{2}}}}\ ,\\
E_{y}\left(  x,y\right)   &  =\frac{k\lambda L}{\sqrt{2}}\frac{\left(
\sqrt{\left(  x-\frac{1}{2}~L\right)  ^{2}+y^{2}}+\sqrt{\left(  x+\frac{1}%
{2}~L\right)  ^{2}+y^{2}}\right)  y}{\sqrt{\left(  \frac{1}{4}~L^{2}%
+x^{2}+y^{2}\right)  ^{2}-L^{2}x^{2}}\sqrt{x^{2}+y^{2}-\frac{1}{4}L^{2}%
+\sqrt{\left(  \frac{1}{4}~L^{2}+x^{2}+y^{2}\right)  ^{2}-L^{2}x^{2}}}}\ .
\end{align}
Although the geometric features of the electric field may not be transparent
from these expressions, nevertheless the fact that the equipotentials are
ellipsoidal allows one to immediately visualize the direction of
$\overrightarrow{E}$\ as normal to those surfaces of constant $\Phi$.

To confirm the charge density along the line, we again use $\int%
\overrightarrow{\nabla}\cdot\overrightarrow{E}~dxdy=2\pi k\int\rho~dxdy$, and
integrate over a horizontal rectangle containing an infinitesimal portion of
the $x$-axis, as the height of the rectangle is taken to zero. \ The crucial
features here are $s-L\rightarrow\left\vert 2x-L\right\vert >0$ as
$y\rightarrow0$ for $\left\vert x\right\vert >L/2$ (away from the segment),
but $s-L\rightarrow0$ as $y\rightarrow0$ for $-L/2<x<L/2$ (on the segment).
\ More precisely, as points on the segment are approached transversely,
\begin{gather}
s=\left(  \frac{1}{2}~L-x\right)  \sqrt{1+\frac{y^{2}}{\left(  \frac{1}%
{2}~L-x\right)  ^{2}}}+\left(  \frac{1}{2}~L+x\right)  \sqrt{1+\frac{y^{2}%
}{\left(  \frac{1}{2}~L+x\right)  ^{2}}}\nonumber\\
\underset{y\rightarrow0}{\sim}L+\frac{1}{2}\left(  \frac{1}{\frac{1}{2}%
~L-x}+\frac{1}{\frac{1}{2}~L+x}\right)  y^{2}+O\left(  y^{4}\right)
=L+\frac{1}{2}~L\left(  \frac{1}{\frac{1}{4}~L^{2}-x^{2}}\right)
y^{2}+O\left(  y^{4}\right)  \ .
\end{gather}
Thus for $-L/2<x<L/2$,
\begin{equation}
\lim_{y\rightarrow0}\left(  \frac{y}{\sqrt{s^{2}-L^{2}}}\right)  =\frac
{1}{\sqrt{2L}}\lim_{y\rightarrow0}\frac{y}{\sqrt{s-L}}=\frac{1}{\sqrt{2L}}%
\lim_{y\rightarrow0}\frac{y}{\sqrt{\frac{1}{2}~L\left(  \frac{1}{\frac{1}%
{4}~L^{2}-x^{2}}\right)  y^{2}}}=\frac{1}{L}\sqrt{\frac{1}{4}~L^{2}-x^{2}}\ .
\end{equation}
Combining this with the more obvious $\lim_{y\rightarrow0}\left(  r_{+}%
r_{-}\right)  =\frac{1}{4}~L^{2}-x^{2}$ for $-L/2<x<L/2$ gives%
\begin{equation}
2\pi k~\lambda\left(  x\right)  =2\lim_{y\rightarrow0}E_{y}\left(  x,y\right)
=\lim_{y\rightarrow0}\left(  \frac{2k\lambda Ls}{r_{+}r_{-}}\frac{y}%
{\sqrt{s^{2}-L^{2}}}\right)  =\frac{2k\lambda L}{\sqrt{\frac{1}{4}~L^{2}%
-x^{2}}}\times\left\{
\begin{array}
[c]{cl}%
0 & \text{ \ \ if \ \ }x>L/2\\
1 & \text{ \ \ if \ \ }-L/2<x<L/2\\
0 & \text{ \ \ if \ \ }x<-L/2
\end{array}
\right.  \ . \label{NonUniformCharge2DDerivation}%
\end{equation}
That is to say, the charge/length between $-L/2$ and $L/2$ is%
\begin{equation}
\lambda\left(  x\right)  =\frac{2\lambda L}{\pi\sqrt{L^{2}-4x^{2}}}\ ,
\label{NonUniformCharge2D}%
\end{equation}
as anticipated above in (\ref{Lambda(x)2D}). \ Elsewhere, the charge density
vanishes, again as follows from $\nabla^{2}\Phi_{\text{line}}\left(
\overrightarrow{r}\right)  =0$ for all points not coincident with the segment.

Once more it is instructive to make a vector field plot of $\left(
E_{x}\left(  x,y\right)  ,E_{y}\left(  x,y\right)  \right)  $, especially near
the charged segment. \ The fact that the segment itself is an equipotential is
evident graphically since the electric field lines are perpendicular to the
segment as the $x$-axis is approached for $-L/2<x<L/2$. \ Here are vector
field plots for $L=2$.%
%TCIMACRO{\FRAME{dtbpF}{6.3286in}{6.2985in}{0pt}{}{}%
%{eurojphyslinesegmentsv2__11.pdf}{\special{ language "Scientific Word";
%type "GRAPHIC";  maintain-aspect-ratio TRUE;  display "USEDEF";
%valid_file "F";  width 6.3286in;  height 6.2985in;  depth 0pt;
%original-width 6.4209in;  original-height 6.3907in;  cropleft "0";
%croptop "1";  cropright "1";  cropbottom "0";
%filename 'EuroJPhysLineSegmentsV2__11.pdf';file-properties "XNPEU";}} }%
%BeginExpansion
\begin{center}
\includegraphics[
height=6.2985in,
width=6.3286in
]%
{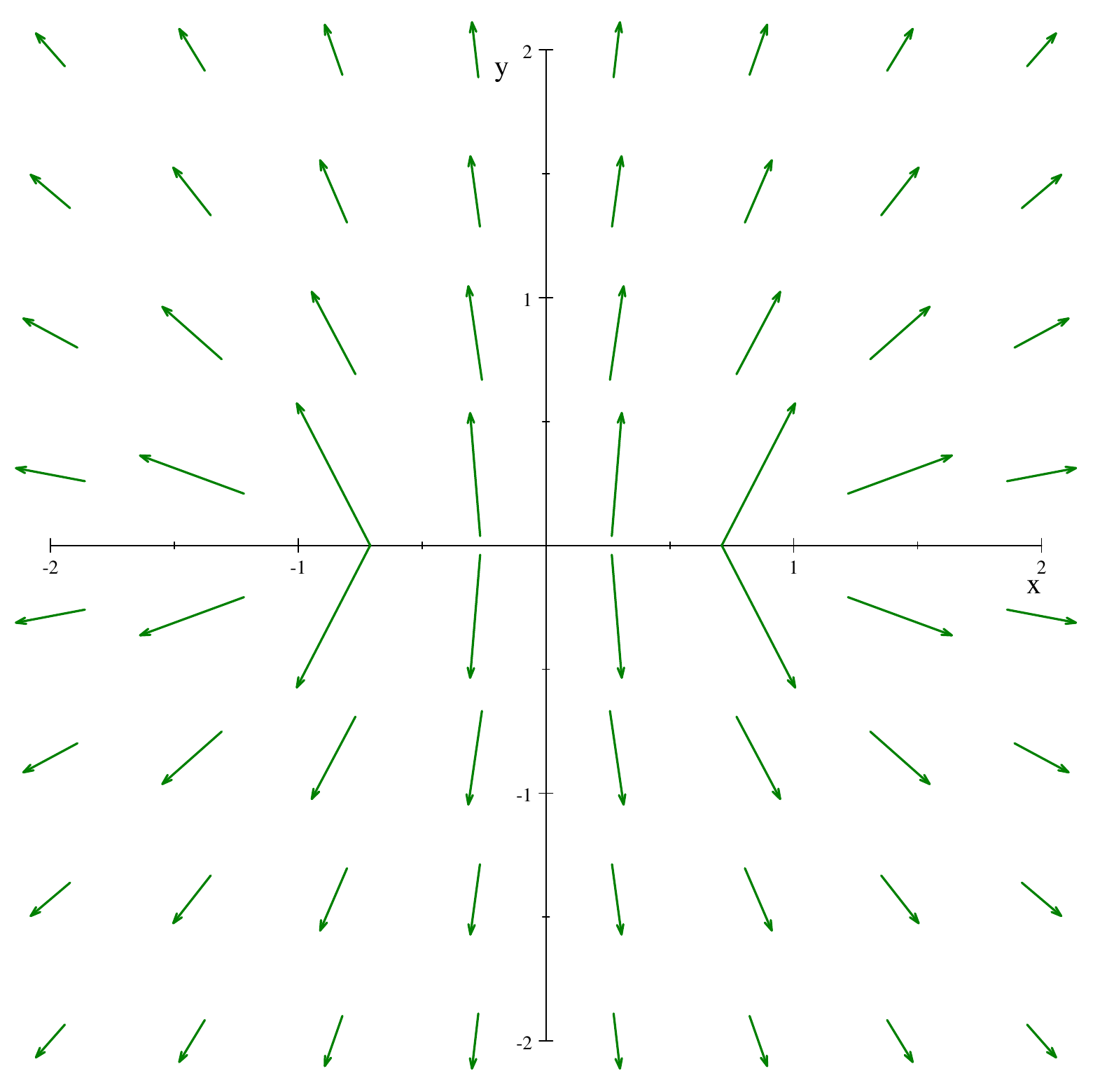}%
\end{center}
%EndExpansion%
%TCIMACRO{\FRAME{dtbpFU}{6.7348in}{2.0667in}{0pt}{\Qcb{Figures 11 \& 12: Vector
%field plots of $\protect\overrightarrow{E}$ for a non-uniformly charged line
%segment situated between $x=-1$ and $x=+1$. \ The field is evaluated at the
%center of each arrow.}}{}{eurojphyslinesegmentsv2__12.pdf}%
%{\special{ language "Scientific Word";  type "GRAPHIC";
%maintain-aspect-ratio TRUE;  display "USEDEF";  valid_file "F";
%width 6.7348in;  height 2.0667in;  depth 0pt;  original-width 6.8351in;
%original-height 2.0773in;  cropleft "0";  croptop "1";  cropright "1";
%cropbottom "0";
%filename 'EuroJPhysLineSegmentsV2__12.pdf';file-properties "XNPEU";}} }%
%BeginExpansion
\begin{center}
\includegraphics[
height=2.0667in,
width=6.7348in
]%
{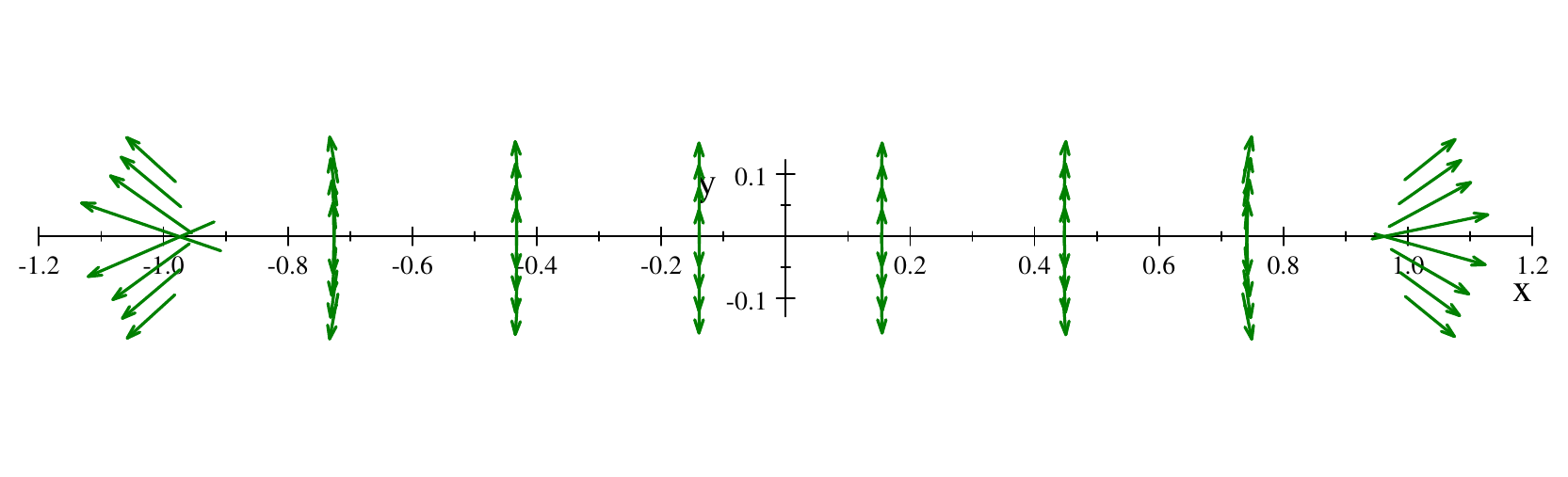}%
\\
Figures 11 \& 12: Vector field plots of $\protect\overrightarrow{E}$ for a
non-uniformly charged line segment situated between $x=-1$ and $x=+1$. \ The
field is evaluated at the center of each arrow.
\end{center}
%EndExpansion

As previously stressed, the geometry of the ellipsoidal equipotentials ensures
that the electric field at any observation point always has a direction that
bisects the angle formed by the two lines from the end points of the segment
to the observation point \cite{Footnote2}. \ This result is also manifest in
an integral expression for $\overrightarrow{E}$ that follows from linear
superposition of the field contributions from infinitesimal $\lambda\left(
x\right)  dx$ charges along the segment, upon choosing an appropriate
integration variable. \ That is,%
\begin{equation}
\overrightarrow{E}\left(  x,y\right)  =\frac{k\lambda L\sqrt{\sin\theta
_{+}\sin\theta_{-}}}{\pi y}\int_{\frac{1}{2}\left(  \theta_{+}-\theta
_{-}\right)  }^{\frac{1}{2}\left(  \theta_{-}-\theta_{+}\right)  }%
\frac{\widehat{n}\left(  \psi\right)  }{\sqrt{\sin\left(  \psi+\frac{1}%
{2}\left(  \theta_{-}-\theta_{+}\right)  \right)  \sin\left(  \frac{1}%
{2}\left(  \theta_{-}-\theta_{+}\right)  -\psi\right)  }}~d\psi\ .
\label{EFieldPunchline}%
\end{equation}
The unit vector $\widehat{n}\left(  \psi\right)  $ points from the
infinitesimal charge on the segment to the observation point, with angle
$\theta=\psi+\frac{1}{2}\left(  \theta_{+}+\theta_{-}\right)  $ measured from
the $x$-axis in the usual counterclockwise sense on the $xy$-plane. \ Thus
$\cos\theta=\widehat{x}\cdot\widehat{n}\left(  \psi\right)  $.
\ Correspondingly,\ $\theta_{\pm}$ are the angles from the ends of the line
segment to the observation point, as given by $\cos\theta_{\pm}=\widehat{x}%
\cdot\widehat{r}_{\pm}$. \ Since the integration over $\psi$ weights
$\widehat{n}\left(  \psi\right)  $ by an \emph{even} function of $\psi$, and
the range of integration is \emph{symmetric} about $\psi=0$, it follows that
the resulting direction of $\overrightarrow{E}\left(  x,y\right)  $ will be
proportional to $\widehat{n}\left(  \psi=0\right)  =\left(  \widehat{r}%
_{+}+\widehat{r}_{-}\right)  /\left\vert \widehat{r}_{+}+\widehat{r}%
_{-}\right\vert $ which points in direction $\frac{1}{2}\left(  \theta
_{+}+\theta_{-}\right)  $. \ That is to say, \
\begin{equation}
\widehat{x}\cdot\overrightarrow{E}\left(  x,y\right)  =\left\vert
\overrightarrow{E}\left(  x,y\right)  \right\vert \cos\left(  \frac{\theta
_{+}+\theta_{-}}{2}\right)  \ ,\ \ \widehat{y}\cdot\overrightarrow{E}\left(
x,y\right)  =\left\vert \overrightarrow{E}\left(  x,y\right)  \right\vert
\sin\left(  \frac{\theta_{+}+\theta_{-}}{2}\right)  \ .
\end{equation}
The change of variables needed to obtain (\ref{EFieldPunchline}) will be
discussed more fully below, for non-uniformly charged segments giving rise to
ellipsoidal equipotentials in any number of dimensions.

A direct graphical comparison of the uniformly charged segment and the
non-uniformly charged segment in 2D is obtained by superimposing their
equipotentials in a true orthogonal projection of the $\Phi$ surface contours
onto the $xy$-plane.%
%TCIMACRO{\FRAME{dtbpFU}{6.294in}{4.1874in}{0pt}{\Qcb{Figure 13: \ 2D
%equipotential contours for a uniformly charged segment, with constant
%$\lambda$ for $-1\leq x\leq1$, for $\Phi=1.5$,\ $1.0$,\ $0.5$,\ $0.0$, $-0.5$,
%and $-1.0$, as inner to outer black curves, and ellipsoidal contours for a
%coincident non-uniformly charged segment, for $\Phi=1.0$, $0.0$, and $-1.0$,
%as inner to outer red curves.}}{}{eurojphyslinesegmentsv2__13.pdf}%
%{\special{ language "Scientific Word";  type "GRAPHIC";
%maintain-aspect-ratio TRUE;  display "USEDEF";  valid_file "F";
%width 6.294in;  height 4.1874in;  depth 0pt;  original-width 6.3854in;
%original-height 4.2389in;  cropleft "0";  croptop "1";  cropright "1";
%cropbottom "0";
%filename 'EuroJPhysLineSegmentsV2__13.pdf';file-properties "XNPEU";}} }%
%BeginExpansion
\begin{center}
\includegraphics[
height=4.1874in,
width=6.294in
]%
{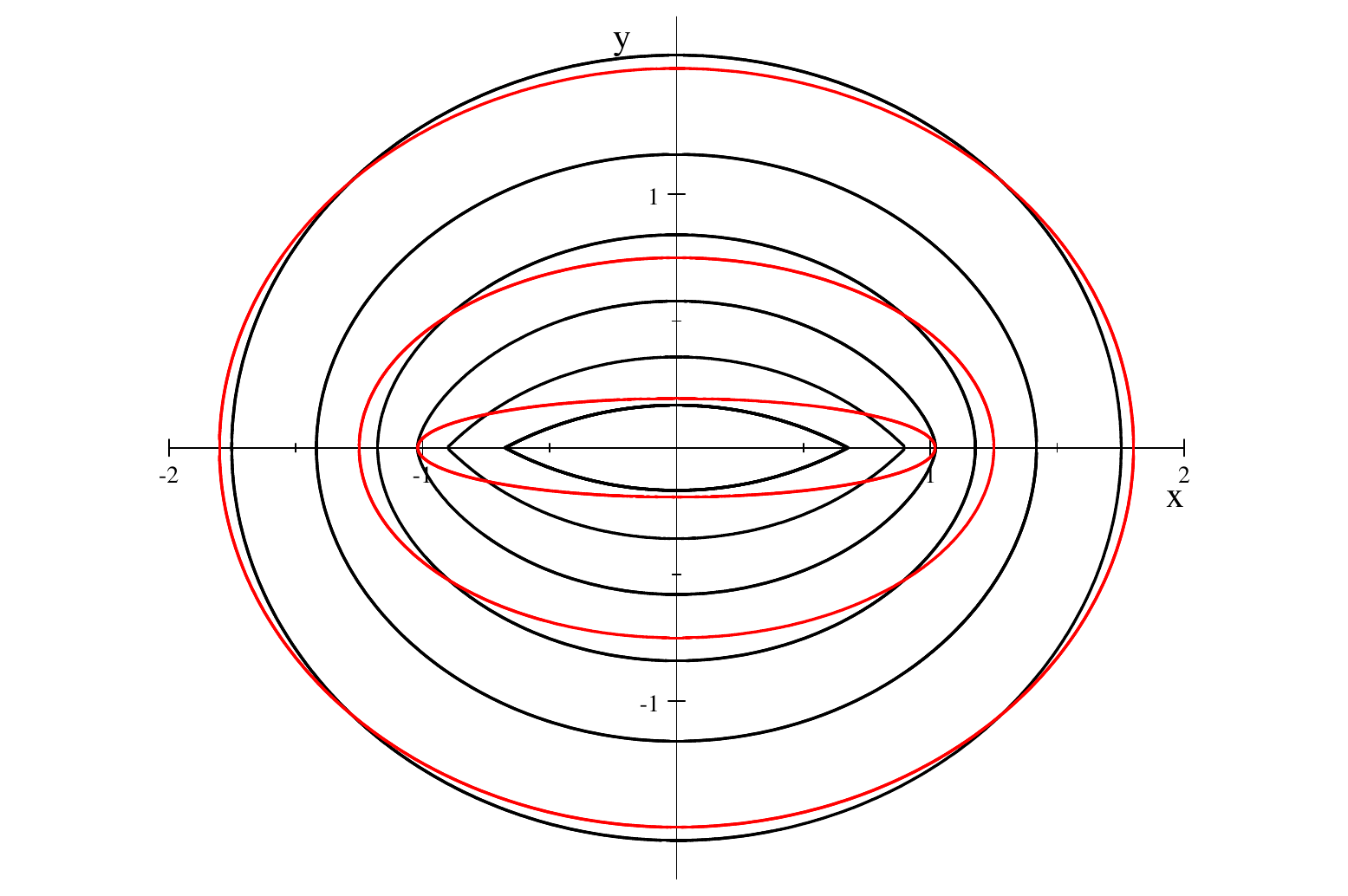}%
\\
Figure 13: \ 2D equipotential contours for a uniformly charged segment, with
constant $\lambda$ for $-1\leq x\leq1$, for $\Phi=1.5$,\ $1.0$,\ $0.5$%
,\ $0.0$, $-0.5$, and $-1.0$, as inner to outer black curves, and ellipsoidal
contours for a coincident non-uniformly charged segment, for $\Phi=1.0$,
$0.0$, and $-1.0$, as inner to outer red curves.
\end{center}
%EndExpansion

For large distances from the line segments, whether uniformly charged or
otherwise, the potential approaches that of a point charge as given in
(\ref{2DPointCharge}). \ So asymptotically both sets of contours become
circles. \ Regarding this, recall the total charge on either segment under
consideration is the same, namely, $Q=\lambda L$. \ Thus for large distances
from the segment the equipotentials in Figure 13 will coalesce, although it is
perhaps surprising how rapidly this occurs. \ In Figure 13 the equipotential
contours for both the uniform and non-uniform charge distributions, for the
same value of $\Phi$, are very nearly coincident if $r\gtrapprox2L$ --- the
exact locations of the two sets of contours never differ by more than a few
percent if $r\gtrapprox2L$ --- well before the contours reach their asymptotic
circular form.

\section{Point charge in $D$ dimensions}

For a point charge $Q$ located at the origin of coordinates, in $D>2$
dimensions, the scalar potential at an observation point $\overrightarrow{r}%
$\ is hypothesized to be \cite{Sommerfeld6}
\begin{equation}
\Phi_{\text{point}}\left(  D,\overrightarrow{r}\right)  =\frac{kQ}{r^{D-2}}\ ,
\end{equation}
where $k$ is the $D$-dimensional analogue of Coulomb's constant. \ Integrating
the radial gradient of $\Phi_{\text{point}}$ over the surface of a sphere
fixes the normalization, by Gauss' law.%
\begin{gather}
\overrightarrow{E}_{\text{point}}\left(  \overrightarrow{r}\right)
=-\overrightarrow{\nabla}\Phi_{\text{point}}\left(  D,\overrightarrow{r}%
\right)  =-\widehat{r}~\partial_{r}\Phi_{\text{point}}\left(
D,\overrightarrow{r}\right)  =\frac{\left(  D-2\right)  kQ~\widehat{r}%
}{r^{D-1}}\ ,\\
\int_{r\leq R}\overrightarrow{\nabla}\cdot\overrightarrow{E}\left(
\overrightarrow{r}\right)  ~d^{D}r=\int_{S_{D-1}}\overrightarrow{E}\left(
R\widehat{r}\right)  \cdot\widehat{r}~R^{D-1}d\Omega=\left(  D-2\right)
\Omega_{D}~kQ\ , \label{IntegralGaussLawInD}%
\end{gather}
where the total solid angle (i.e. the area of the \emph{unit radius} sphere,
$S_{D-1}$, embedded in $D$ dimensions) is given by
\begin{equation}
\Omega_{D}=\int_{S_{D-1}}d\Omega=\frac{2\pi^{D/2}}{\Gamma\left(  D/2\right)
}\ .
\end{equation}
For example, $\Omega_{1}=2$, $\Omega_{2}=2\pi$, $\Omega_{3}=4\pi$, $\Omega
_{4}=2\pi^{2}$, etc.

To put it differently, in terms of a Dirac delta in $D$-dimensions,%
\begin{align}
\overrightarrow{\nabla}\cdot\overrightarrow{E}_{\text{point}}\left(
\overrightarrow{r}\right)   &  =\left(  D-2\right)  \Omega_{D}~kQ~\delta
^{D}\left(  \overrightarrow{r}\right)  \ ,\label{DifferentialGaussLawInD}\\
\nabla^{2}\left(  \frac{1}{r^{D-2}}\right)   &  =-\left(  D-2\right)
\Omega_{D}~\delta^{D}\left(  \overrightarrow{r}\right)  \ .
\end{align}
And in fact, this gives the correct result even for $D=2$, by taking a limit: \ 

$\nabla^{2}\left(  1-\left(  D-2\right)  \ln r+O\left(  \left(  D-2\right)
^{2}\right)  \right)  \underset{D\rightarrow2}{\sim}-\left(  D-2\right)
\Omega_{2}~\delta^{2}\left(  \overrightarrow{r}\right)  $, hence $\nabla
^{2}\ln r=2\pi\delta^{2}\left(  \overrightarrow{r}\right)  $. \ So for $D=2$
the point particle potential is logarithmic, as previously noted in Section 2.

Perhaps the simplest convention would be to set
\[
Q=\int\overrightarrow{E}\left(  \overrightarrow{r}\right)  \cdot
\widehat{r}~r^{D-1}d\Omega\ ,\ \ \ \nabla^{2}\Phi_{\text{point}}\left(
D,\overrightarrow{r}\right)  =-Q~\delta^{D}\left(  \overrightarrow{r}\right)
\ ,
\]
which would require, for $D>2$,%
\[
k=\frac{1}{\left(  D-2\right)  \Omega_{D}}=\frac{\Gamma\left(  D/2\right)
}{\left(  D-2\right)  2\pi^{D/2}}\ .
\]
This is singular at $D=2$ where the potential is a logarithm, not a power.
\ In that case the corresponding choice would be $k=\frac{1}{2\pi}$, so
\[
\Phi_{\text{point}}\left(  D=2,\overrightarrow{r}\right)  =-\frac{Q}{2\pi}~\ln
r\ ,
\]
and again $\nabla^{2}\Phi_{\text{point}}\left(  D=2,\overrightarrow{r}\right)
=-Q~\delta^{2}\left(  \overrightarrow{r}\right)  $.

\newpage

\section{Uniformly charged line segments for all dimensions $D>2$}

The line of charge and observation point are shown in red in the following
Figure.%
%TCIMACRO{\FRAME{dtbpFU}{5.7574in}{4.7959in}{0pt}{\Qcb{Figure 14: \ Segment
%coordinates.}}{}{eurojphyslinesegmentsv2__14.pdf}%
%{\special{ language "Scientific Word";  type "GRAPHIC";
%maintain-aspect-ratio TRUE;  display "USEDEF";  valid_file "F";
%width 5.7574in;  height 4.7959in;  depth 0pt;  original-width 6.3135in;
%original-height 5.2536in;  cropleft "0";  croptop "1";  cropright "1";
%cropbottom "0";
%filename 'EuroJPhysLineSegmentsV2__14.pdf';file-properties "XNPEU";}} }%
%BeginExpansion
\begin{center}
\includegraphics[
height=4.7959in,
width=5.7574in
]%
{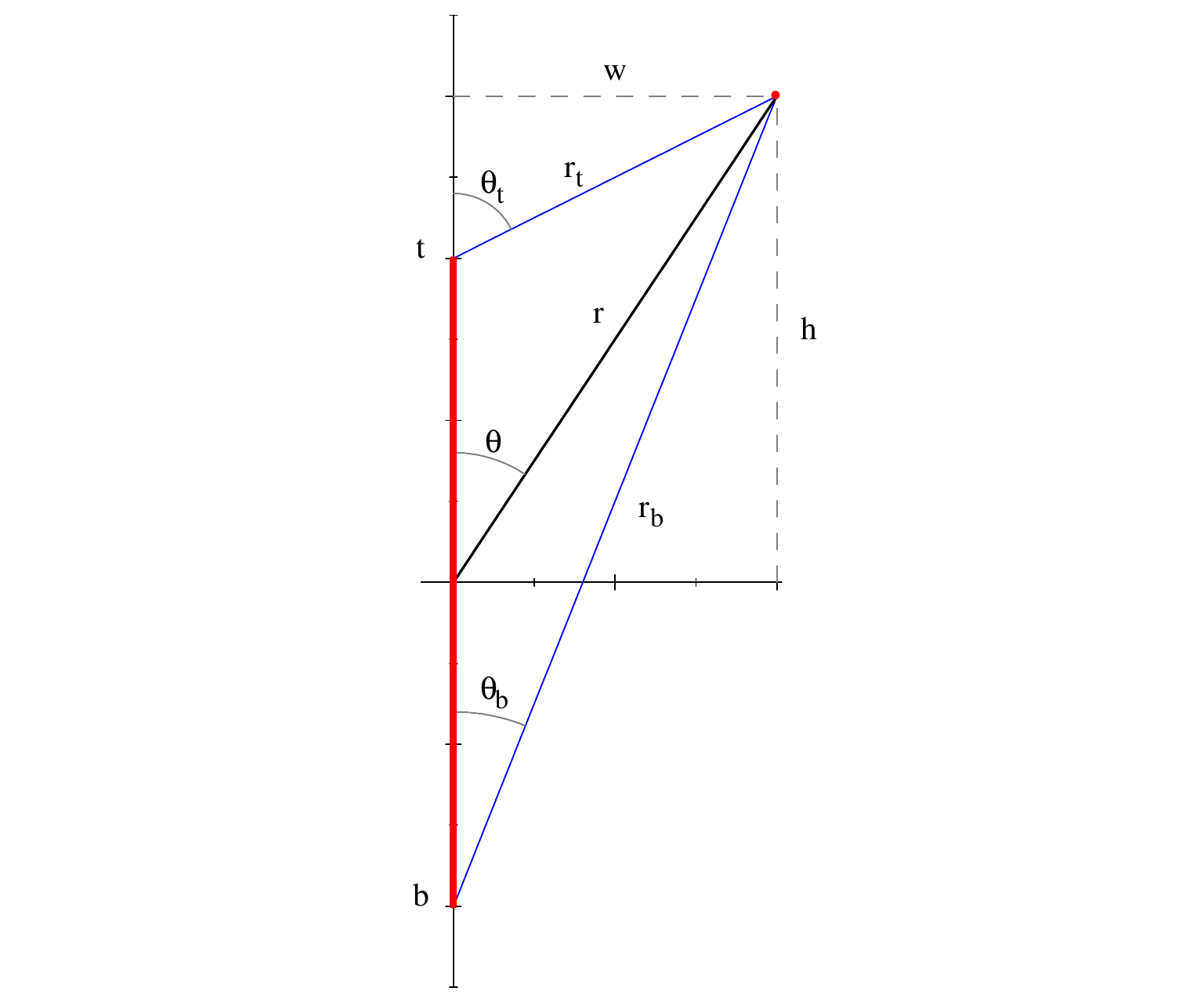}%
\\
Figure 14: \ Segment coordinates.
\end{center}
%EndExpansion
From the Figure ($+$ and $-$ in the formulas correspond, respectively, to the
bottom point \textquotedblleft b\textquotedblright\ and the top point
\textquotedblleft t\textquotedblright\ in the Figure)
\begin{subequations}
\begin{align}
r_{\text{t}} &  =\sqrt{\frac{1}{4}L^{2}-Lr\cos\theta+r^{2}}%
\ ,\ \ \ r_{\text{b}}=\sqrt{\frac{1}{4}L^{2}+Lr\cos\theta+r^{2}}%
\ ,\label{Trig}\\
\sin\theta_{\text{b,t}} &  =\frac{r\sin\theta}{\sqrt{\frac{1}{4}L^{2}\pm
Lr\cos\theta+r^{2}}}\ ,\ \ \ \cos\theta_{\text{b,t}}=\frac{r\cos\theta\pm
L/2}{\sqrt{\frac{1}{4}L^{2}\pm Lr\cos\theta+r^{2}}}\ ,\\
\frac{\sin\theta_{\text{b}}}{r_{\text{t}}} &  =\frac{\sin\theta_{\text{t}}%
}{r_{\text{b}}}=\frac{\sin\left(  \theta_{\text{t}}-\theta_{\text{b}}\right)
}{L}\ ,\\
\theta_{\text{t}}-\theta_{\text{b}} &  =\arcsin\left(  \frac{Lr\sin\theta
}{\sqrt{\left(  \frac{1}{4}L^{2}+r^{2}\right)  ^{2}-L^{2}r^{2}\cos^{2}\theta}%
}\right)  \ .\label{arcsin}%
\end{align}
The relations in the third line above follow from the Law of Sines.$\ \ $Note
that the last expression must be used with care since $\arcsin$\ is
multi-valued. \ In particular, if $r<L/2$, then as the segment is approached
it is always true that $\theta_{\text{t}}-\theta_{\text{b}}\rightarrow\pi$, no
matter how the segment is approached. \ 

More generally, when $r$ decreases to cross the surface of the sphere for
which the segment is a diameter, the value of $\arcsin$\ increases through
$\pi/2$. \ This follows from $\left.  \sqrt{\left(  \frac{1}{4}L^{2}%
+r^{2}\right)  ^{2}-L^{2}r^{2}\cos^{2}\theta}\right\vert _{r=L/2}=\frac{1}%
{2}L^{2}\sin\theta$ so that the argument of the $\arcsin$ in (\ref{arcsin}) is
just $1$ for $r=L/2$, and $\theta_{\text{t}}-\theta_{\text{b}}=\pi/2$ at this
radius. \ Reducing $r$ below $L/2$ increases $\theta_{\text{t}}-\theta
_{\text{b}}$ above $\pi/2$, i.e. $\arcsin$ has moved onto another branch of
the function. This change of branch can be explicitly taken into account
through the use of Heaviside step functions, $\Theta$, to write
\end{subequations}
\begin{equation}
\theta_{\text{t}}-\theta_{\text{b}}=\pi~\Theta\left(  \frac{L}{2}-r\right)
+\left(  \Theta\left(  r-\frac{L}{2}\right)  -\Theta\left(  \frac{L}%
{2}-r\right)  \right)  \arcsin\left(  \frac{Lr\sin\theta}{\sqrt{\left(
\frac{1}{4}L^{2}+r^{2}\right)  ^{2}-L^{2}r^{2}\cos^{2}\theta}}\right)  \ ,
\label{ArcsinMulti}%
\end{equation}
where the $\arcsin$ in this expression is the principal branch of the function.

Assuming linear superposition for the potential, a uniformly charged line
segment as shown above in Figure 14, of length $L$, centered on the origin,
with top (t) and bottom (b) ends at $\theta=0$ (i.e. $z=L/2$) and $\theta=\pi$
(i.e. $z=-L/2$), will produce in $D>2$ dimensions a potential $\Phi
_{\text{line}}$ given by \cite{Footnote3}
\begin{equation}
\Phi_{\text{line}}\left(  D,r,\theta\right)  =\frac{k\lambda}{\left(
r\sin\theta\right)  ^{D-3}}\int_{\theta_{\text{b}}}^{\theta_{\text{t}}}\left(
\sin\vartheta\right)  ^{D-4}~d\vartheta\ . \label{PhiD}%
\end{equation}
Here $\lambda$ is the constant charge/length on the line segment, and
$\theta_{\text{t,b}}$ are the polar angles for vectors from the top and bottom
endpoints of the line segment to the observation point $\left(  r,\theta
\right)  $, as in Figure 14.

The potential has no dependence on the additional $D-2$ angles needed to
specify the location of a point in $D$ dimensions using spherical polar
coordinates. \ That is to say, in $D$ dimensions the equipotentials are always
higher dimensional surfaces of revolution about the line segment. \ The
corresponding electric field is given by%
\begin{equation}
\overrightarrow{E}_{\text{line}}\left(  \overrightarrow{r}\right)
=-\overrightarrow{\nabla}\Phi_{\text{line}}=-\widehat{r}~\partial_{r}%
\Phi_{\text{line}}-\frac{1}{r}~\widehat{\theta}~\partial_{\theta}%
\Phi_{\text{line}}\ , \label{LineE}%
\end{equation}
and it depends manifestly on $r$ and $\theta$. \ The only dependence of the
electric field on the additional $D-2$ angles in $D$ dimensional spherical
polar coordinates is carried by the unit vectors $\widehat{r}$\ and
$\widehat{\theta}$.

To see that (\ref{PhiD}) is correct, we need only sum the potential
contributions for infinitesimal pieces of the line segment with charge
\begin{equation}
dQ=\lambda dz\ ,
\end{equation}
located on the vertical axis of Figure 14 at position $z$, and at a distance
$\mathcal{\ell}\left(  z\right)  $ from the observation point $\left(
r,\theta\right)  $ as given by
\begin{equation}
\mathcal{\ell}\left(  z\right)  =\sqrt{r^{2}+z^{2}-2rz\cos\theta}\ .
\end{equation}
But writing $z=h-w\cot\vartheta$ and $\mathcal{\ell}\left(  z\right)
=\frac{w}{\sin\vartheta}$, for fixed $h$ and $w$, we have $dz=-w~d\cot
\vartheta=\frac{w}{\sin^{2}\vartheta}~d\vartheta$, and
\begin{equation}
\int_{-L/2}^{L/2}\frac{dz}{\left(  \mathcal{\ell}\left(  z\right)  \right)
^{D-2}}=\frac{1}{w^{D-3}}\int_{\theta_{\text{b}}}^{\theta_{\text{t}}}\left(
\sin\vartheta\right)  ^{D-4}~d\vartheta\ .
\end{equation}
On the other hand, $w=r\sin\theta$. \ Hence the result (\ref{PhiD}).

For instance, in the special case $D=3$ the potential of the uniformly charged
segment is
\begin{equation}
\Phi_{\text{line}}\left(  D=3,r,\theta\right)  =k\lambda\int_{\theta
_{\text{b}}}^{\theta_{\text{t}}}\frac{1}{\sin\vartheta}~d\vartheta
=k\lambda\left.  \ln\left(  \frac{\sin\vartheta}{1+\cos\vartheta}\right)
\right\vert _{\vartheta=\theta_{\text{b}}}^{\vartheta=\theta_{\text{t}}}\ .
\end{equation}
Note that the potential is infinite for all points on the charged segment
itself, for which points $\theta_{\text{b}}=0$ and $\theta_{\text{t}}=\pi$.
\ Be that as it may, after a bit of algebra $\Phi_{\text{line}}\left(
D=3\right)  $ can be reduced to
\begin{equation}
\Phi_{\text{line}}\left(  D=3,r,\theta\right)  =k\lambda\ln\left(  \frac
{s+L}{s-L}\right)  \ ,
\end{equation}
where
\begin{equation}
s=r_{\text{t}}+r_{\text{b}} \label{s}%
\end{equation}
is the sum of the distances from the end points of the segment to the
observation point. \ Thus equipotentials in this case are given by constant
$s$, and as is common knowledge, this defines an ellipsoid of revolution about
the line segment with focal points $t$ and $b$. \ A view\ from below the
potential surface again clearly shows the equipotentials are ellipses.%
%TCIMACRO{\FRAME{dtbpFU}{6.2355in}{4.1458in}{0pt}{\Qcb{Figure 15: $\ \frac
%{1}{k\lambda}~\Phi_{\text{line}}\left(  D=3,r,\theta\right)  =\ln\left(
%\frac{\sqrt{1-2z+r^{2}}+\sqrt{1+2z+r^{2}}+2}{\sqrt{1-2z+r^{2}}+\sqrt
%{1+2z+r^{2}}-2}\right)  $ for $L=2$, where $z=r\cos\theta$ and $x=r\sin\theta
%$.}}{}{eurojphyslinesegmentsv2__15.pdf}{\special{ language "Scientific Word";
%type "GRAPHIC";  maintain-aspect-ratio TRUE;  display "USEDEF";
%valid_file "F";  width 6.2355in;  height 4.1458in;  depth 0pt;
%original-width 6.326in;  original-height 4.1981in;  cropleft "0";
%croptop "1";  cropright "1";  cropbottom "0";
%filename 'EuroJPhysLineSegmentsV2__15.pdf';file-properties "XNPEU";}} }%
%BeginExpansion
\begin{center}
\includegraphics[
height=4.1458in,
width=6.2355in
]%
{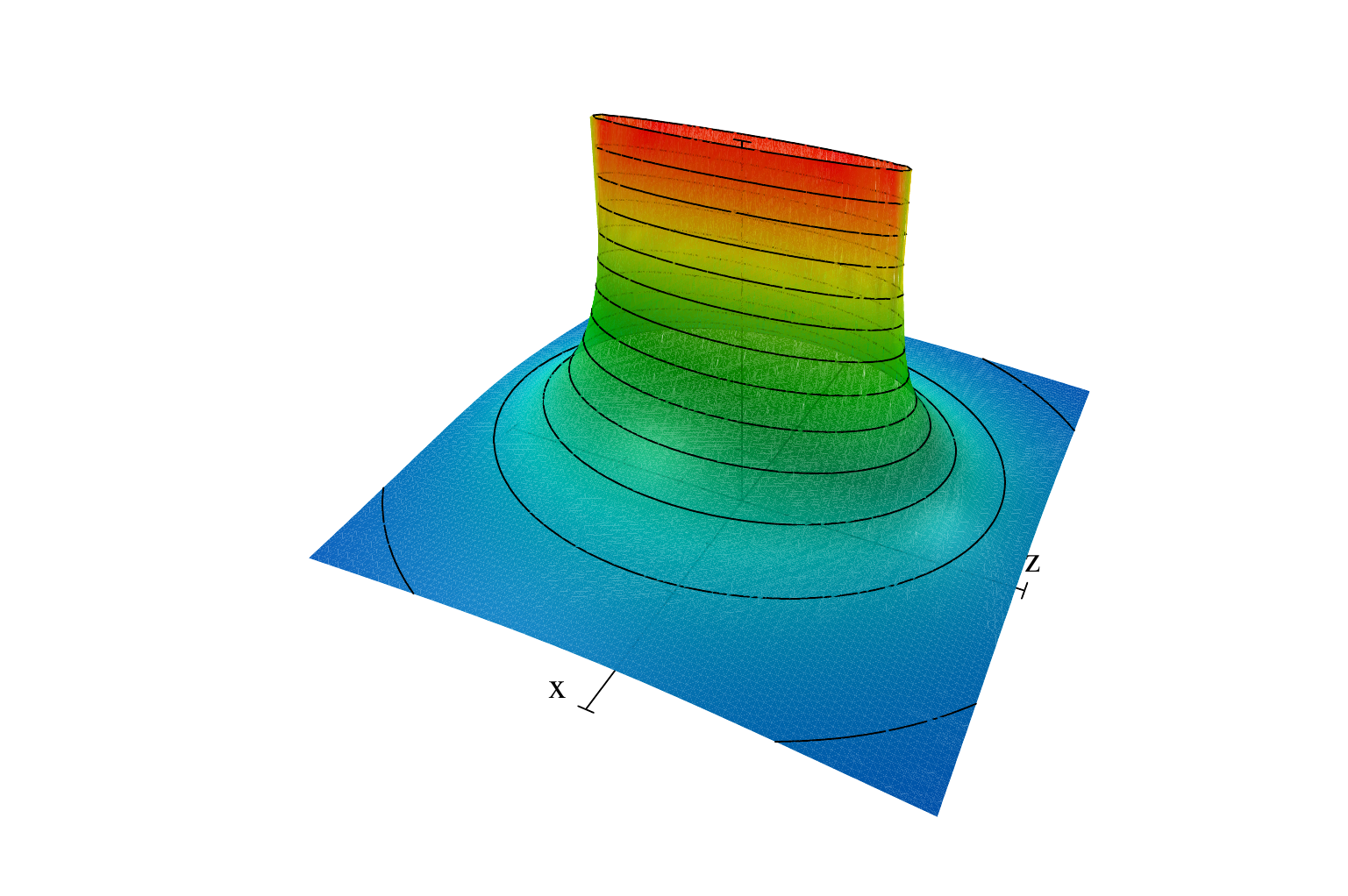}%
\\
Figure 15: $\ \frac{1}{k\lambda}~\Phi_{\text{line}}\left(  D=3,r,\theta
\right)  =\ln\left(  \frac{\sqrt{1-2z+r^{2}}+\sqrt{1+2z+r^{2}}+2}%
{\sqrt{1-2z+r^{2}}+\sqrt{1+2z+r^{2}}-2}\right)  $ for $L=2$, where
$z=r\cos\theta$ and $x=r\sin\theta$.
\end{center}
%EndExpansion%
%TCIMACRO{\FRAME{dtbpFU}{6.2222in}{4.1316in}{0pt}{\Qcb{Figure 16: \ Contours of
%constant $\frac{1}{k\lambda}~\Phi_{\text{line}}\left(  D=3,r,\theta\right)
%=\ln\left(  \frac{\sqrt{1-2z+r^{2}}+\sqrt{1+2z+r^{2}}+2}{\sqrt{1-2z+r^{2}%
%}+\sqrt{1+2z+r^{2}}-2}\right)  $ for $L=2$, plotted versus $z=r\cos\theta$
%(vertical axis) and $x=r\sin\theta$ (horizontal axis).}}{}%
%{eurojphyslinesegmentsv2__16.pdf}{\special{ language "Scientific Word";
%type "GRAPHIC";  maintain-aspect-ratio TRUE;  display "USEDEF";
%valid_file "F";  width 6.2222in;  height 4.1316in;  depth 0pt;
%original-width 6.3135in;  original-height 4.183in;  cropleft "0";
%croptop "1";  cropright "1";  cropbottom "0";
%filename 'EuroJPhysLineSegmentsV2__16.pdf';file-properties "XNPEU";}} }%
%BeginExpansion
\begin{center}
\includegraphics[
height=4.1316in,
width=6.2222in
]%
{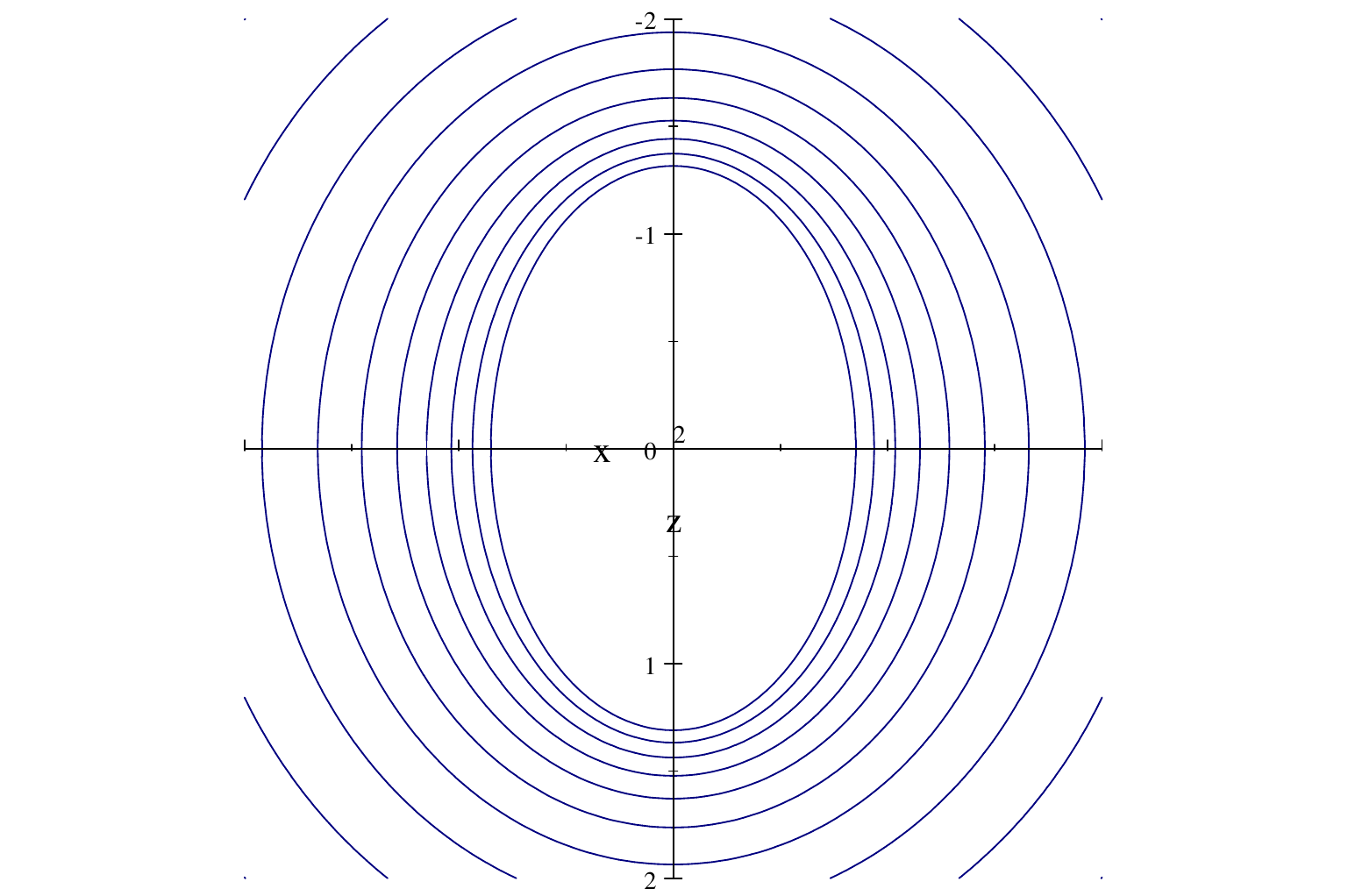}%
\\
Figure 16: \ Contours of constant $\frac{1}{k\lambda}~\Phi_{\text{line}%
}\left(  D=3,r,\theta\right)  =\ln\left(  \frac{\sqrt{1-2z+r^{2}}%
+\sqrt{1+2z+r^{2}}+2}{\sqrt{1-2z+r^{2}}+\sqrt{1+2z+r^{2}}-2}\right)  $ for
$L=2$, plotted versus $z=r\cos\theta$ (vertical axis) and $x=r\sin\theta$
(horizontal axis).
\end{center}
%EndExpansion

For other dimensions, however, the geometrical shapes of the equipotentials
are not so easily discerned. \ For $D>3$, it remains to evaluate the angular
integral in (\ref{PhiD}). \ Define the indefinite integral $I\left(  D\right)
=\int\left(  \sin\vartheta\right)  ^{D-4}~d\vartheta$, and compute%
\begin{equation}%
\begin{array}
[c]{ll}%
I\left(  4\right)  =\vartheta\ , & I\left(  5\right)  =-\cos\vartheta\ ,\\
& \\
I\left(  6\right)  =\frac{1}{2}\vartheta-\frac{1}{2}\cos\vartheta\sin
\vartheta\ , & I\left(  7\right)  =\frac{1}{3}\cos^{3}\vartheta-\cos
\vartheta\ ,\\
& \\
I\left(  8\right)  =\frac{3}{8}\vartheta+\left(  \frac{1}{4}\cos^{3}%
\vartheta-\frac{5}{8}\cos\vartheta\right)  \sin\vartheta\ ,\ \ \ \ \  &
I\left(  9\right)  =-\frac{1}{5}\cos^{5}\vartheta+\frac{2}{3}\cos^{3}%
\vartheta-\cos\vartheta\ ,
\end{array}
\end{equation}
etc. \ For odd $D\geq5$ the integral is always a polynomial of order $D-4$ in
$\cos\vartheta$, while for even $D\geq4$ the integral always has a term linear
in $\vartheta$\ plus, for $D\geq6$, a term with $\sin\vartheta$ multiplying a
polynomial of order $D-5$ in $\cos\vartheta$. \ For any $D$ we therefore
obtain the angular integral in (\ref{PhiD})\ in terms of $\left(
\theta_{\text{t}}-\theta_{\text{b}}\right)  $, $\sin\theta_{\text{t}}$,
$\sin\theta_{\text{b}}$, $\cos\theta_{\text{t}}$, and $\cos\theta_{\text{b}}$.
\ These quantities may then be expressed in terms of $\theta$ and $r$ upon
using the relations in (\ref{Trig}-\ref{arcsin}).

For example, in $D=4$ the potential of the uniformly charged segment is%
\begin{equation}
\Phi_{\text{line}}\left(  D=4,r,\theta\right)  =\frac{k\lambda}{r\sin\theta
}\left(  \theta_{\text{t}}-\theta_{\text{b}}\right)  \ .
\end{equation}
Unlike $\Phi_{\text{line}}\left(  D=3\right)  $, the potential $\Phi
_{\text{line}}\left(  D=4\right)  $ is \emph{not} a function solely of the
variable $s$ as defined in (\ref{s}). \ Consequently, the equipotentials are
\emph{not} ellipsoidal for this four dimensional example. \ Indeed, a plot now
shows that the equipotentials are not ellipsoidal, especially for points close
to the uniform line of charge, although the shape of the potential surface is
similar to that for $D=3$.%
%TCIMACRO{\FRAME{dtbpFU}{6.2222in}{4.1316in}{0pt}{\Qcb{Figure 17: $\ \frac
%{1}{k\lambda}~\Phi_{\text{line}}\left(  D=4,r,\theta\right)  =\frac
%{1}{\left\vert x\right\vert }\arcsin\left(  \frac{2\left\vert x\right\vert
%}{\sqrt{\left(  1+r^{2}\right)  ^{2}-4z^{2}}}\right)  $ for $L=2$, where
%$z=r\cos\theta$ and $x=r\sin\theta$.}}{}{eurojphyslinesegmentsv2__17.pdf}%
%{\special{ language "Scientific Word";  type "GRAPHIC";
%maintain-aspect-ratio TRUE;  display "USEDEF";  valid_file "F";
%width 6.2222in;  height 4.1316in;  depth 0pt;  original-width 6.3135in;
%original-height 4.183in;  cropleft "0";  croptop "1";  cropright "1";
%cropbottom "0";
%filename 'EuroJPhysLineSegmentsV2__17.pdf';file-properties "XNPEU";}} }%
%BeginExpansion
\begin{center}
\includegraphics[
height=4.1316in,
width=6.2222in
]%
{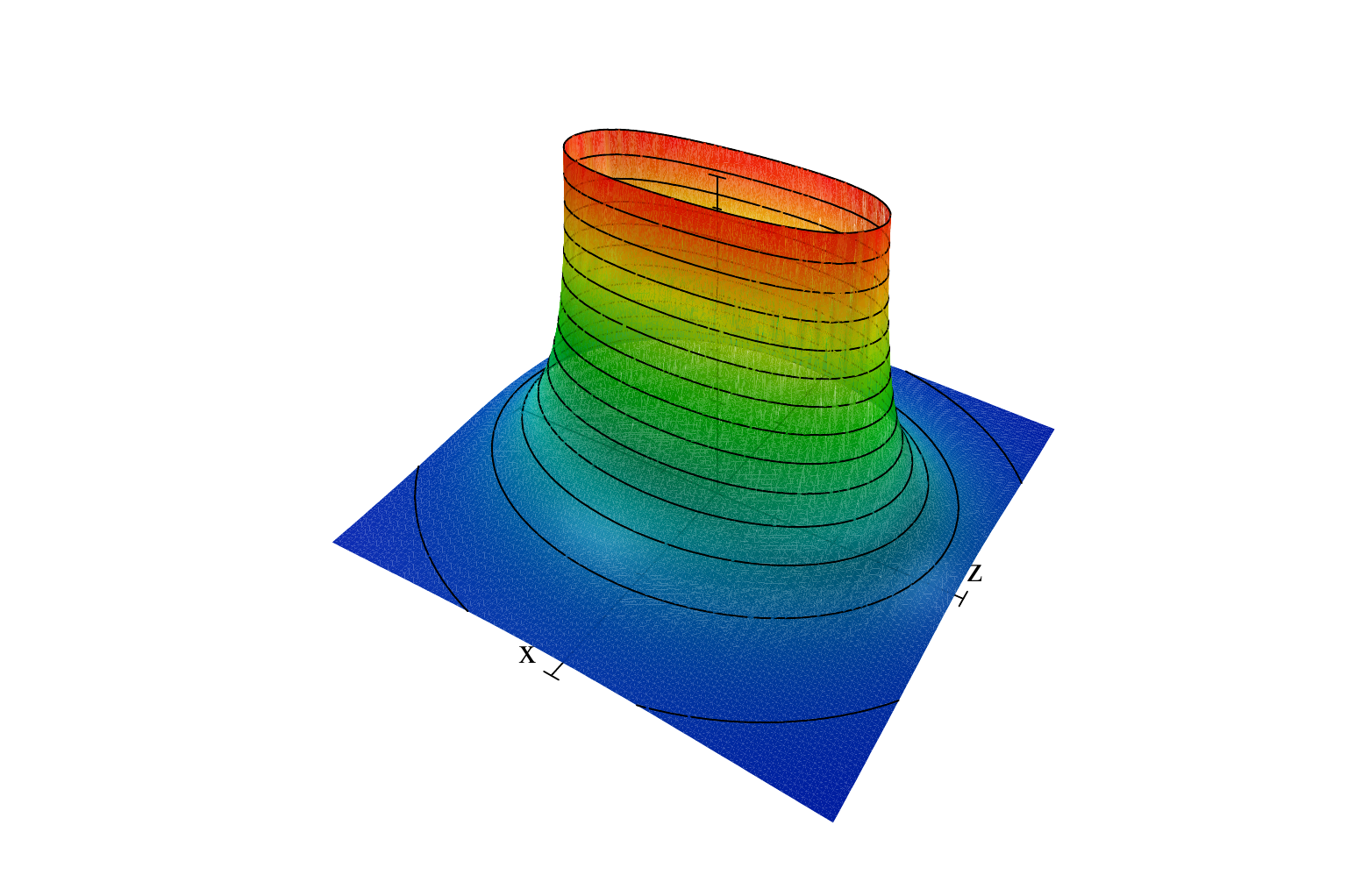}%
\\
Figure 17: $\ \frac{1}{k\lambda}~\Phi_{\text{line}}\left(  D=4,r,\theta
\right)  =\frac{1}{\left\vert x\right\vert }\arcsin\left(  \frac{2\left\vert
x\right\vert }{\sqrt{\left(  1+r^{2}\right)  ^{2}-4z^{2}}}\right)  $ for
$L=2$, where $z=r\cos\theta$ and $x=r\sin\theta$.
\end{center}
%EndExpansion
Note that $\Phi_{\text{line}}$ is infinite for all points on the segment.
\ Also note, to obtain the correct shape of the potential surface it is
important to interpret $\arcsin$ in the formula for $\Phi_{\text{line}}$ as a
multi-valued function, as given explicitly by (\ref{ArcsinMulti}). \ 

A view from below the potential surface shows that the potential contours are
\emph{not} ellipses in this case, although the difference is somewhat subtle.
\ This next graph should be compared closely to the ellipsoidal case for $D=4$
as presented below in Section 5.%
%TCIMACRO{\FRAME{dtbpFU}{6.2222in}{4.1316in}{0pt}{\Qcb{Figure 18: \ Contours of
%constant $\frac{1}{k\lambda}~\Phi_{\text{line}}\left(  D=4,r,\theta\right)
%=\frac{1}{\left\vert x\right\vert }\arcsin\left(  \frac{2\left\vert
%x\right\vert }{\sqrt{\left(  1+r^{2}\right)  ^{2}-4z^{2}}}\right)  $ for
%$L=2$.}}{}{eurojphyslinesegmentsv2__18.pdf}%
%{\special{ language "Scientific Word";  type "GRAPHIC";
%maintain-aspect-ratio TRUE;  display "USEDEF";  valid_file "F";
%width 6.2222in;  height 4.1316in;  depth 0pt;  original-width 6.3135in;
%original-height 4.183in;  cropleft "0";  croptop "1";  cropright "1";
%cropbottom "0";
%filename 'EuroJPhysLineSegmentsV2__18.pdf';file-properties "XNPEU";}} }%
%BeginExpansion
\begin{center}
\includegraphics[
height=4.1316in,
width=6.2222in
]%
{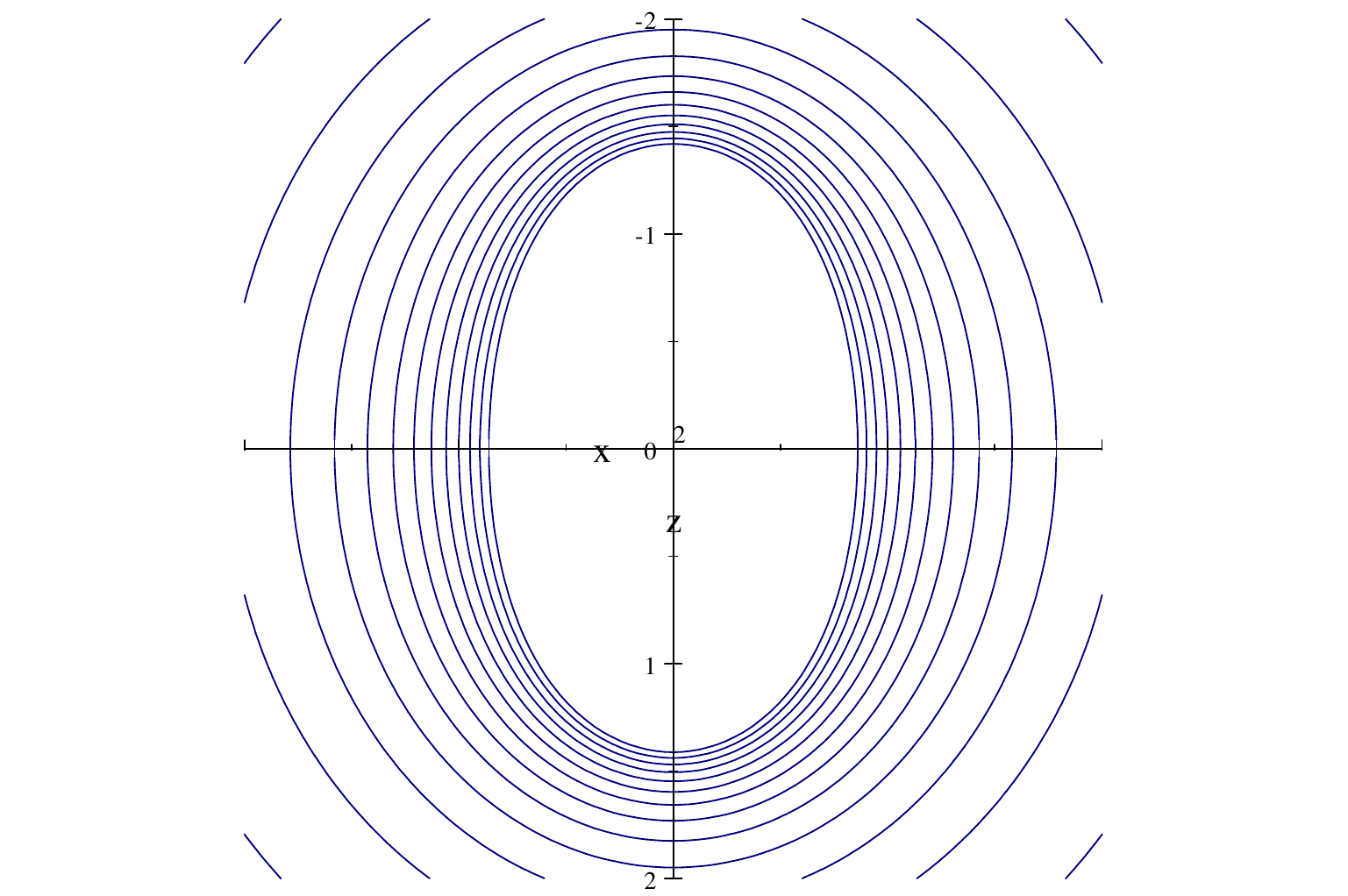}%
\\
Figure 18: \ Contours of constant $\frac{1}{k\lambda}~\Phi_{\text{line}%
}\left(  D=4,r,\theta\right)  =\frac{1}{\left\vert x\right\vert }%
\arcsin\left(  \frac{2\left\vert x\right\vert }{\sqrt{\left(  1+r^{2}\right)
^{2}-4z^{2}}}\right)  $ for $L=2$.
\end{center}
%EndExpansion

The electric field is given by (\ref{LineE}) and in this case has $r$ and
$\theta$ components%
\begin{align}
-\partial_{r}\Phi_{\text{line}}\left(  D=4\right)   &  =-\frac{k\lambda}%
{r\sin\theta}~\partial_{r}\left(  \theta_{\text{t}}-\theta_{\text{b}}\right)
+\frac{1}{r}~\Phi_{\text{line}}\left(  D=4\right)  \ ,\label{UniformEr}\\
-\frac{1}{r}~\partial_{\theta}\Phi_{\text{line}}\left(  D=4\right)   &
=-\frac{k\lambda}{r^{2}\sin\theta}~\partial_{\theta}\left(  \theta_{\text{t}%
}-\theta_{\text{b}}\right)  +\frac{\cot\theta}{r}~\Phi_{\text{line}}\left(
D=4\right)  \ . \label{UniformEtheta}%
\end{align}
All the complication in these expressions lies in the derivatives
$\partial_{r}\left(  \theta_{\text{t}}-\theta_{\text{b}}\right)  $\ and
$\partial_{\theta}\left(  \theta_{\text{t}}-\theta_{\text{b}}\right)  $ whose
geometrical significance is not yet easy to visualize. \ On the other hand,
the last terms involving $\frac{1}{r}~\Phi_{\text{line}}$ in (\ref{UniformEr})
and (\ref{UniformEtheta}) have simple geometrical interpretations since they
define a vector that is perpendicular to the axis of the segment and points
away from that axis for positive $\lambda$\ (i.e. in the direction of
$\widehat{\rho}=\widehat{r}\sin\theta+\widehat{\theta}\cos\theta$ if we were
to use cylindrical coordinates).

In any case, it is worthwhile to compute $\partial_{r}\left(  \theta
_{\text{t}}-\theta_{\text{b}}\right)  $\ and $\partial_{\theta}\left(
\theta_{\text{t}}-\theta_{\text{b}}\right)  $ since these derivatives appear
for all even $D\geq4$ and always have the same form. \ After some algebra we
find
\begin{align}
\partial_{r}\left(  \theta_{\text{t}}-\theta_{\text{b}}\right)   &
=\frac{4\left(  L^{2}+4r^{2}\right)  L\sin\theta}{\left(  L^{2}+4r^{2}\right)
^{2}-16L^{2}r^{2}\cos^{2}\theta}~\operatorname*{sgn}\left(  L^{2}%
-4r^{2}\right)  \ ,\\
\frac{1}{r}~\partial_{\theta}\left(  \theta_{\text{t}}-\theta_{\text{b}%
}\right)   &  =\frac{4\left(  L^{2}-4r^{2}\right)  L\cos\theta}{\left(
L^{2}+4r^{2}\right)  ^{2}-16L^{2}r^{2}\cos^{2}\theta}~\operatorname*{sgn}%
\left(  L^{2}-4r^{2}\right)  \ .
\end{align}
Note the difference in sign for $r^{2}\gtrless\frac{1}{4}L^{2}$. \ Once again
$\widehat{\rho}=\widehat{r}\sin\theta+\widehat{\theta}\cos\theta$, so the bulk
of the contributions from these derivatives again gives a vector that is
perpendicular to the axis of the segment. \ But there remains a contribution
that is entirely in the $\widehat{r}$ direction if $L^{2}+4r^{2} $ is written
as $L^{2}-4r^{2}+8r^{2}$, or else entirely in the $\widehat{\theta}$ direction
if $L^{2}-4r^{2}$ is written as $L^{2}+4r^{2}-8r^{2}$. \ Choosing the first of
these options gives%
\begin{align}
\overrightarrow{E}_{\text{line}}\left(  D=4,\overrightarrow{r}\right)   &
=\left(  \frac{1}{r}~\Phi_{\text{line}}\left(  D=4,r,\theta\right)
-\frac{4Lk\lambda}{r}\frac{\left\vert L^{2}-4r^{2}\right\vert }{\left(
L^{2}+4r^{2}\right)  ^{2}-16L^{2}r^{2}\cos^{2}\theta}\right)  \left(
\widehat{r}+\widehat{\theta}\cot\theta\right) \nonumber\\
&  -\left(  \frac{32k\lambda rL}{\left(  L^{2}+4r^{2}\right)  ^{2}%
-16L^{2}r^{2}\cos^{2}\theta}~\operatorname*{sgn}\left(  L^{2}-4r^{2}\right)
\right)  \widehat{r}%
\end{align}
To compare to the three dimensional ellipsoidal case, it suffices to express
$\widehat{r}$ and $\widehat{\theta}$ in terms of $\widehat{r}_{\text{t}}$ and
$\widehat{r}_{\text{b}}$. \ Although the latter two unit vectors are not
orthogonal, they are independent except at $\theta=0$ and $\theta=\pi$, and at
those particular angles either one of $\widehat{r}_{\text{t}}$ and
$\widehat{r}_{\text{b}}$\ will suffice to give the direction of the electric
field, since $\overrightarrow{E}_{\text{line}}\left(  \text{on the
}z\text{-axis}\right)  \propto\widehat{z}$. \ We find%
\begin{align}
\widehat{r}  &  =\frac{1}{2r}\left(  \overrightarrow{r}_{\text{t}%
}+\overrightarrow{r}_{\text{b}}\right) \\
&  =\left(  \frac{1}{4r}\sqrt{L^{2}-4Lr\cos\theta+4r^{2}}\right)
~\widehat{r}_{\text{t}}+\left(  \frac{1}{4r}\sqrt{L^{2}+4Lr\cos\theta+4r^{2}%
}\right)  ~\widehat{r}_{\text{b}}\ ,\nonumber
\end{align}%
\begin{align}
\widehat{\theta}  &  =\frac{1}{Lr^{2}\sin\theta}\left(  \left(
\overrightarrow{r}\cdot\overrightarrow{r}_{\text{b}}\right)
\overrightarrow{r}_{\text{t}}-\left(  \overrightarrow{r}\cdot
\overrightarrow{r}_{\text{t}}\right)  \overrightarrow{r}_{\text{b}}\right) \\
&  =\frac{2r+L\cos\theta}{4Lr\sin\theta}\sqrt{L^{2}-4Lr\cos\theta+4r^{2}%
}~\widehat{r}_{\text{t}}-\frac{2r-L\cos\theta}{4Lr\sin\theta}\sqrt
{L^{2}+4Lr\cos\theta+4r^{2}}~\widehat{r}_{\text{b}}\ .\nonumber
\end{align}

\section{Ellipsoidal equipotentials for all dimensions $D>2$}

Consider the same geometry as in Figure 14, but rather than assuming uniform
charge density on the line segment, suppose the potential produced by the line
segment depends positionally only on $s$ in any number of dimensions.
\ Including some convenient numerical and dimensionful factors, suppose%
\begin{equation}
\Phi\left(  \overrightarrow{r}\right)  =k\lambda L~\left(  \frac{2}{L}\right)
^{D-2}V\left(  s\right)  \ ,\ \ \ s=r_{\text{t}}+r_{\text{b}}\ .
\label{EllipsoidalAnyD}%
\end{equation}
In this case, equipotentials are ellipsoidal by assumption, since the
positional dependence is only on $s$. \ The only issue is to determine
$V\left(  s\right)  $.

If the only charge present is on the line segment, then for other points the
potential function must be harmonic, $\nabla^{2}\Phi\left(  \overrightarrow{r}%
\right)  =0$. \ So, using $\overrightarrow{\nabla}r_{\text{t}}%
=\overrightarrow{r_{\text{t}}}/r_{\text{t}}$ and $\overrightarrow{\nabla
}r_{\text{b}}=\overrightarrow{r_{\text{b}}}/r_{\text{b}}$ (see Eqns(5)-(7) in
\cite{TSvK}\ for more details), we compute%
\begin{equation}
\nabla^{2}V\left(  s\right)  =\frac{1}{r_{\text{t}}r_{\text{b}}}\left(
\left(  D-1\right)  sV^{\prime}\left(  s\right)  +\left(  s^{2}-L^{2}\right)
V^{\prime\prime}\left(  s\right)  \right)  \ .
\end{equation}
The potential will then be harmonic for points not lying on the segment if and
only if%
\begin{equation}
\left(  L^{2}-s^{2}\right)  V^{\prime\prime}\left(  s\right)  +\left(
1-D\right)  sV^{\prime}\left(  s\right)  =0
\end{equation}
for $s>L$. \ For general $D$ \emph{the relevant solution} of this second
order, ordinary differential equation, is given by a Gauss hypergeometric
function represented by the standard series around $s=\infty$, namely,%
\begin{gather}
V\left(  s\right)  =\left(  \frac{s}{L}\right)  ^{2-D}~\left.  _{2}%
F_{1}\right.  \left(  \tfrac{1}{2}D-1,\tfrac{1}{2}D-\tfrac{1}{2};\tfrac{1}%
{2}D;L^{2}/s^{2}\right)  \ ,\label{3PointPotential}\\
\underset{s\rightarrow\infty}{\sim}\left(  \frac{L}{s}\right)  ^{D-2}+O\left(
\frac{L^{D}}{s^{D}}\right)  \text{ \ \ for \ \ }D>2\ ,\\
\underset{s\rightarrow L}{\sim}\frac{D-2}{D-3}\left(  \sqrt{\frac{L/2}{s-L}%
}\right)  ^{D-3}\text{ \ \ for \ \ }D>3\ ,
\end{gather}
The various factors in the last line are\ useful to determine the exact
expression for the charge distribution along the segment, as presented below
(see (\ref{ChargeDistribution})). \ Note the large $s$ behavior of $V$ implies%
\begin{equation}
\Phi\left(  \overrightarrow{r}\right)  \underset{r\rightarrow\infty}{\sim
}\frac{k\lambda L}{r^{D-2}}+O\left(  \frac{1}{r^{D}}\right)
\end{equation}
since $s\underset{r\rightarrow\infty}{\sim}2r$. \ The normalization of $V$ and
the $2/L$ factors in (\ref{EllipsoidalAnyD}) were chosen so that the total
charge on the segment is the same as in the uniformly charged case, namely,
$Q=\lambda L$. \ This total charge appears in the limiting form of the
potential for $r\gg L$, i.e. $\lim\limits_{r\rightarrow\infty}r^{D-2}%
\Phi\left(  \overrightarrow{r}\right)  /k=Q$.

Alternatively, closed-form expressions for $V\left(  s\right)  $ in terms of
more elementary functions are sometimes more easily obtained by evaluating the
corresponding hypergeometric series around $s=0$, as given by \cite{Footnote4}%
\begin{equation}
V\left(  s\right)  =A\left(  D\right)  ~\frac{s}{L}~\left.  _{2}F_{1}\right.
\left(  \tfrac{1}{2},\tfrac{1}{2}D-\tfrac{1}{2};\tfrac{3}{2};s^{2}%
/L^{2}\right)  +B\left(  D\right)  \ , \label{Alternate3PointPotential}%
\end{equation}
and then analytically continuing to $s>L$. \ Here $A\left(  D\right)  $ and
$B\left(  D\right)  $ are constants chosen so that the result
(\ref{Alternate3PointPotential}) is real-valued for $s>L$, and so that any
constant term (the trivial harmonic) is eliminated from $V$\ as $s\rightarrow
\infty$.

For integer $D$ the hypergeometric functions that appear in the solutions
(\ref{3PointPotential}) and (\ref{Alternate3PointPotential}) always reduce to
elementary functions. \ For example, in various dimensions the relevant
harmonic functions are:%
\begin{align}
D  &  =1\ ,\ \ \ V\left(  s\right)  =s/L\\
D  &  =2\ ,\ \ \ V\left(  s\right)  =\ln\left(  \frac{2L}{s+\sqrt{s^{2}-L^{2}%
}}\right) \\
&  \underset{s\rightarrow\infty}{\sim}\ln\left(  \frac{L}{s}\right)  +O\left(
\frac{L^{2}}{s^{2}}\right) \nonumber\\
D  &  =3\ ,\ \ \ V\left(  s\right)  =\frac{1}{2}\ln\left(  \frac{s+L}%
{s-L}\right) \\
&  \underset{s\rightarrow\infty}{\sim}\frac{L}{s}+O\left(  \frac{L^{3}}{s^{3}%
}\right) \nonumber\\
D  &  =4\ ,\ \ \ V\left(  s\right)  =2\left(  \frac{s}{\sqrt{s^{2}-L^{2}}%
}-1\right) \\
&  \underset{s\rightarrow\infty}{\sim}\frac{L^{2}}{s^{2}}+O\left(  \frac
{L^{4}}{s^{4}}\right) \nonumber\\
D  &  =5\ ,\ \ \ V\left(  s\right)  =-\frac{3}{4}\left(  \ln\left(  \frac
{s+L}{s-L}\right)  -\frac{2sL}{s^{2}-L^{2}}\right) \\
&  \underset{s\rightarrow\infty}{\sim}\frac{L^{3}}{s^{3}}+O\left(  \frac
{L^{5}}{s^{5}}\right) \nonumber
\end{align}
By construction, the equipotentials are always ellipsoidal for any $D$, except
for the trivial one dimensional case where equipotentials consist of just
pairs of points on the line. \ In contrast to the uniformly charged line
segment for $D\neq3$, if the potential depends only on $s$ then the charged
line itself --- where $s=L$ for all points on the segment --- is \emph{always}
an equipotential, albeit with \emph{infinite} $\Phi$ when $D\geq3$, while for
$D=1$ and $D=2$ the potential along the charged line is \emph{finite}.

For example, consider graphically the case for $D=4$. \ Qualitatively the
potential surface is similar to that for the uniformly charged segment in
three dimensions. \ Potential contours are ellipsoids that surround $-L/2\leq
z\leq L/2$ and \emph{never} intersect the charge segment. \ Also, the top of
the potential surface is just an exact copy of the segment itself, albeit at
an infinite value of $\Phi$. \ That is to say, the charged segment is itself
an equipotential, but in fact the potential is infinite for points on the
segment. \ Again, a view from below the potential surface shows more clearly
the ellipsoidal potential contours.%
%TCIMACRO{\FRAME{dtbpFU}{6.2142in}{4.1262in}{0pt}{\Qcb{Figure 19: $\ \frac
%{1}{k\lambda}~\Phi=\left.  L~\left(  \frac{2}{L}\right)  ^{D-2}V\left(
%s\right)  \right\vert _{D=4}=\frac{8}{L}\left(  \frac{s}{\sqrt{s^{2}-L^{2}}%
%}-1\right)  $ for $L=2$, versus $z=r\cos\theta$ and $x=r\sin\theta$.}}%
%{}{eurojphyslinesegmentsv2__19.pdf}{\special{ language "Scientific Word";
%type "GRAPHIC";  maintain-aspect-ratio TRUE;  display "USEDEF";
%valid_file "F";  width 6.2142in;  height 4.1262in;  depth 0pt;
%original-width 6.3135in;  original-height 4.183in;  cropleft "0";
%croptop "1";  cropright "1";  cropbottom "0";
%filename 'EuroJPhysLineSegmentsV2__19.pdf';file-properties "XNPEU";}} }%
%BeginExpansion
\begin{center}
\includegraphics[
height=4.1262in,
width=6.2142in
]%
{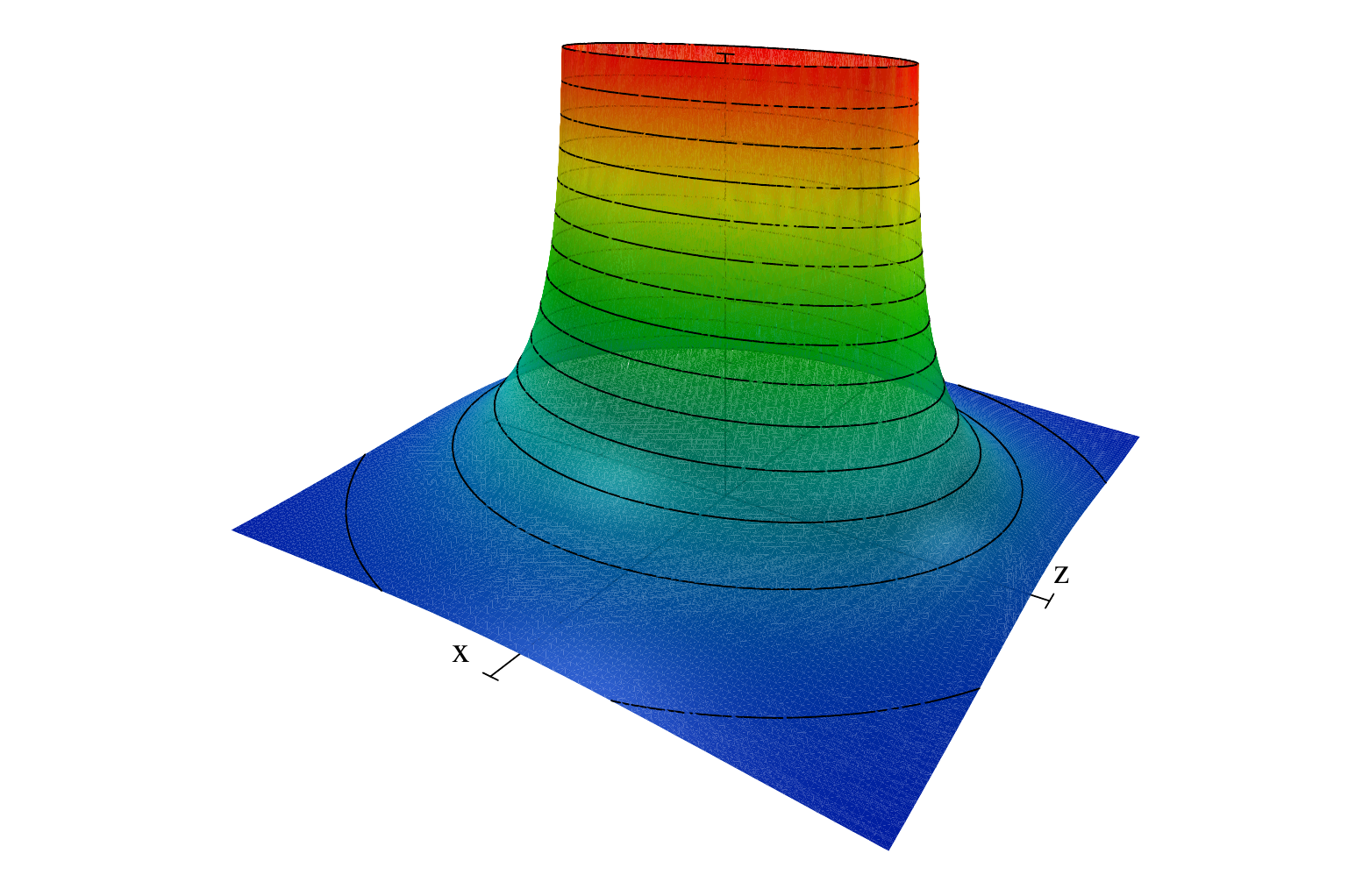}%
\\
Figure 19: $\ \frac{1}{k\lambda}~\Phi=\left.  L~\left(  \frac{2}{L}\right)
^{D-2}V\left(  s\right)  \right\vert _{D=4}=\frac{8}{L}\left(  \frac{s}%
{\sqrt{s^{2}-L^{2}}}-1\right)  $ for $L=2$, versus $z=r\cos\theta$ and
$x=r\sin\theta$.
\end{center}
%EndExpansion%
%TCIMACRO{\FRAME{dtbpFU}{6.2142in}{4.1262in}{0pt}{\Qcb{Figure 20: \ Contours of
%constant $\frac{1}{k\lambda}~\Phi=\left.  L~\left(  \frac{2}{L}\right)
%^{D-2}V\left(  s\right)  \right\vert _{D=4}=\frac{8}{L}\left(  \frac{s}%
%{\sqrt{s^{2}-L^{2}}}-1\right)  $ for $L=2$.}}{}%
%{eurojphyslinesegmentsv2__20.pdf}{\special{ language "Scientific Word";
%type "GRAPHIC";  maintain-aspect-ratio TRUE;  display "USEDEF";
%valid_file "F";  width 6.2142in;  height 4.1262in;  depth 0pt;
%original-width 6.3135in;  original-height 4.183in;  cropleft "0";
%croptop "1";  cropright "1";  cropbottom "0";
%filename 'EuroJPhysLineSegmentsV2__20.pdf';file-properties "XNPEU";}} }%
%BeginExpansion
\begin{center}
\includegraphics[
height=4.1262in,
width=6.2142in
]%
{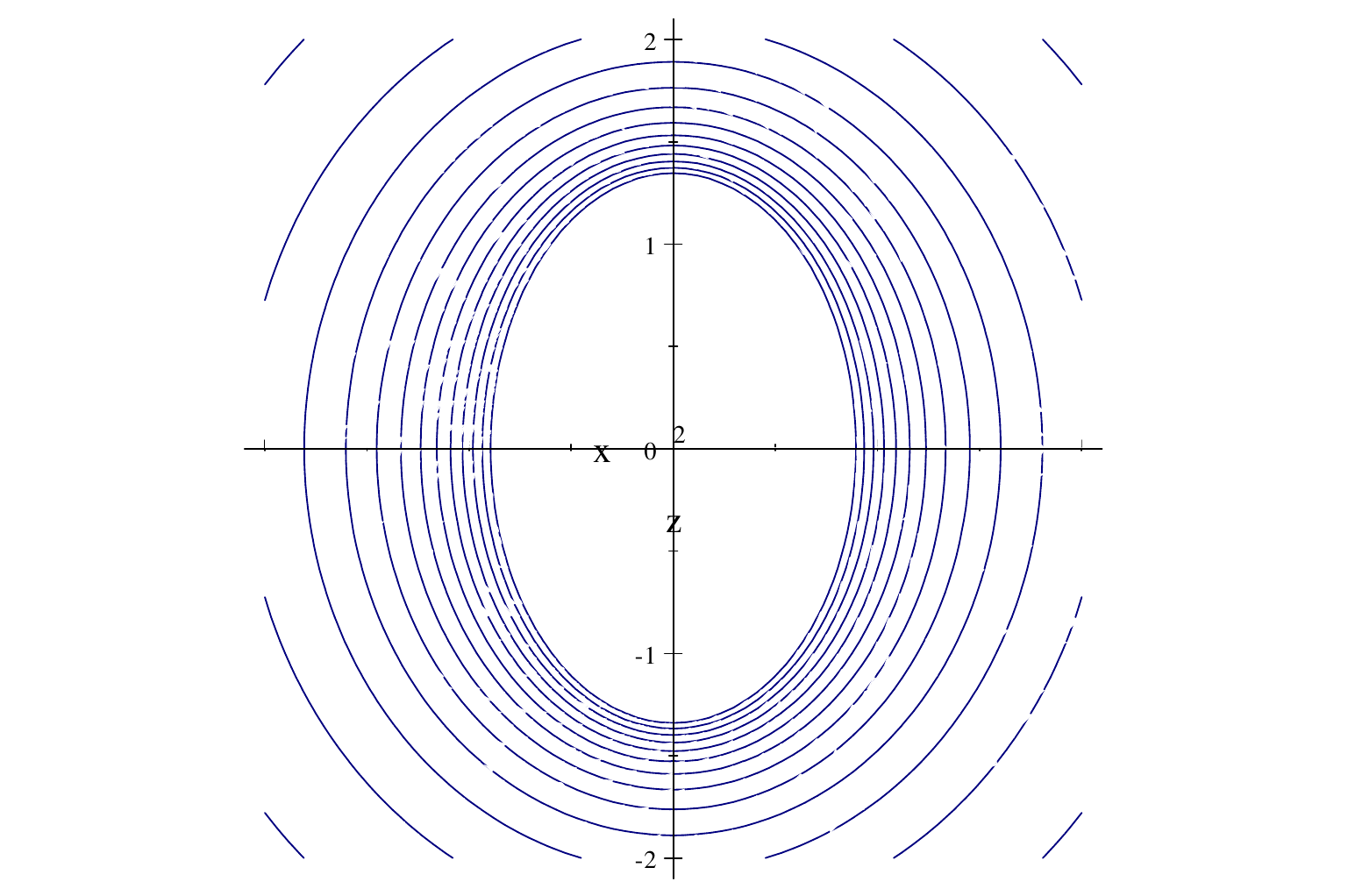}%
\\
Figure 20: \ Contours of constant $\frac{1}{k\lambda}~\Phi=\left.  L~\left(
\frac{2}{L}\right)  ^{D-2}V\left(  s\right)  \right\vert _{D=4}=\frac{8}%
{L}\left(  \frac{s}{\sqrt{s^{2}-L^{2}}}-1\right)  $ for $L=2$.
\end{center}
%EndExpansion

Admittedly, in 4D it takes a discerning eye to see differences in the shape of
the ellipsoidal equipotential surface compared to that for the uniformly
charged segment, if graphs of the two cases are viewed separately. \ But if
viewed side-by-side the difference \emph{is} evident in contour plots for the
potentials. \ Perhaps even more clearly, the difference is highlighted by
plotting both cases in the same graph.%
%TCIMACRO{\FRAME{dtbpFU}{4.528in}{3.0122in}{0pt}{\Qcb{Figure 21: \ Comparison
%of\ $\Phi$ contours in 4D for a uniformly charged segment (black) and a
%non-uniformly charged segment with ellipsoidal equipotentials (red) for
%$L=2$.}}{}{eurojphyslinesegmentsv2__23.pdf}%
%{\special{ language "Scientific Word";  type "GRAPHIC";
%maintain-aspect-ratio TRUE;  display "USEDEF";  valid_file "F";
%width 4.528in;  height 3.0122in;  depth 0pt;  original-width 4.5857in;
%original-height 3.0415in;  cropleft "0";  croptop "1";  cropright "1";
%cropbottom "0";
%filename 'EuroJPhysLineSegmentsV2__23.pdf';file-properties "XNPEU";}} }%
%BeginExpansion
\begin{center}
\includegraphics[
height=3.0122in,
width=4.528in
]%
{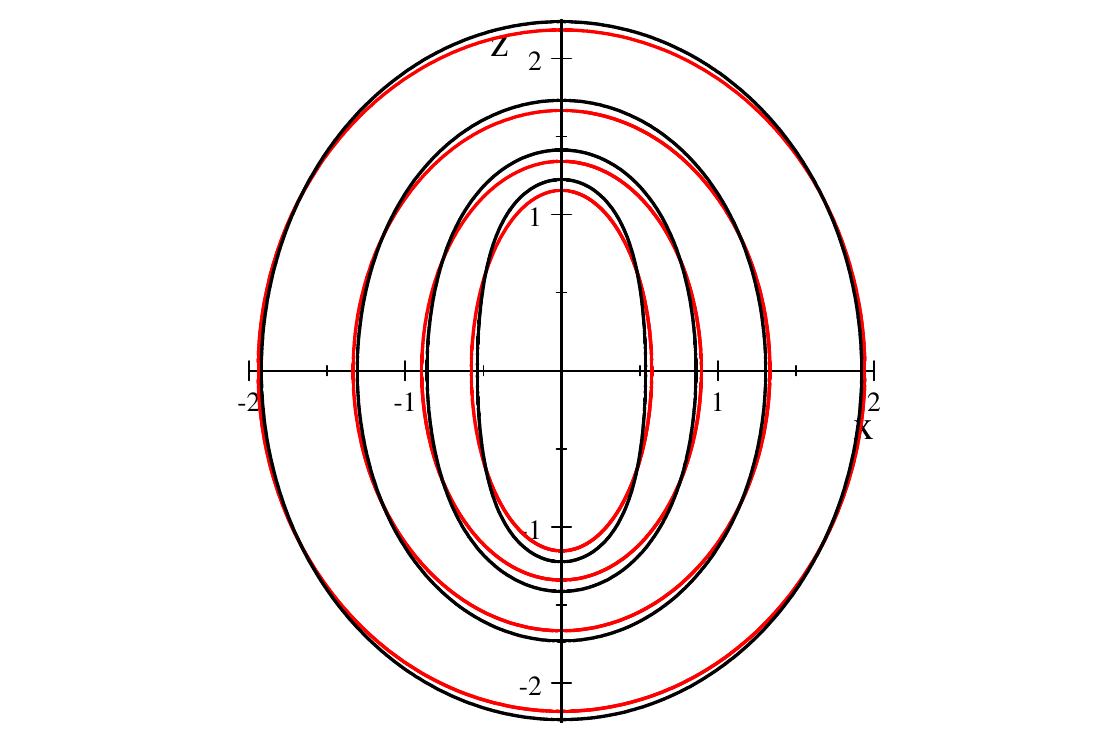}%
\\
Figure 21: \ Comparison of\ $\Phi$ contours in 4D for a uniformly charged
segment (black) and a non-uniformly charged segment with ellipsoidal
equipotentials (red) for $L=2$.
\end{center}
%EndExpansion
Upon doing so, it is also evident that the non-uniform charge distribution
must be greater near the center of the segment rather than at the ends of the
segment, a counter-intuitive feature. \ The ellipsoidal equipotentials are
shorter in the direction of the line segment, and wider transverse to the
segment, than those of the uniformly charged line for the same value of $\Phi
$. \ This is exactly the opposite of what happens in 2D, where the ellipsoidal
equipotentials were longer and narrower than those of the uniformly charged
segment for the same value of $\Phi$. \ 

The non-uniform charge distribution on the segment that produces ellipsoidal
equipotentials can now be determined using the integral form of Gauss' Law in
$D$ dimensions, (\ref{IntegralGaussLawInD}), in complete parallel to the
calculation leading to (\ref{NonUniformCharge2D}) in 2D. \ Here we only give
the results of that calculation.

In $D$ dimensions the charge density for a line segment that produces
ellipsoidal equipotentials is%
\begin{equation}
\lambda_{D}\left(  z\right)  =2\lambda~\frac{\Omega_{D-1}}{\Omega_{D}}\left(
\sqrt{1-\frac{4z^{2}}{L^{2}}}\right)  ^{D-3}\ , \label{ChargeDistribution}%
\end{equation}
where we have used the ratio of total \textquotedblleft solid
angles\textquotedblright\ in $D$ and $D-1$ dimensions,\
\begin{equation}
\frac{\Omega_{D-1}}{\Omega_{D}}=\frac{\Gamma\left(  \frac{1}{2}D\right)
}{\sqrt{\pi}\Gamma\left(  \frac{1}{2}D-\frac{1}{2}\right)  }\ .
\end{equation}
For example, $\left.  \frac{\Gamma\left(  \frac{1}{2}D\right)  }{\sqrt{\pi
}\Gamma\left(  \frac{1}{2}D-\frac{1}{2}\right)  }\right\vert _{D=2}=\frac
{1}{\pi}$ to give $\lambda_{D=2}\left(  z\right)  $ in agreement with
(\ref{Lambda(x)2D}), while $\left.  \frac{\Gamma\left(  \frac{1}{2}D\right)
}{\sqrt{\pi}\Gamma\left(  \frac{1}{2}D-\frac{1}{2}\right)  }\right\vert
_{D=3}=\frac{1}{2}$ to give the uniform distribution, $\lambda_{D=3}\left(
z\right)  =\lambda$. \ These charge densities are all normalized so that the
total charge on the segment is the same for any $D$, namely,%
\begin{equation}
Q=\lambda L=\int_{-L/2}^{L/2}\lambda_{D}\left(  z\right)  dz\ .
\end{equation}
This follows from%
\begin{align}
\int_{-L/2}^{L/2}\lambda_{D}\left(  z\right)  dz  &  =\lambda L~\frac
{\Omega_{D-1}}{\Omega_{D}}\int_{-1}^{1}\left(  1-u^{2}\right)  ^{\frac{D-3}%
{2}}du\ ,\\
\int_{-1}^{1}\left(  1-u^{2}\right)  ^{\frac{D-3}{2}}du  &  =\frac{\sqrt{\pi
}\Gamma\left(  \frac{1}{2}D-\frac{1}{2}\right)  }{\Gamma\left(  \frac{1}%
{2}D\right)  }=\frac{\Omega_{D}}{\Omega_{D-1}}\ .
\end{align}

A geometrical construction to obtain the distribution
(\ref{ChargeDistribution}) is to consider a uniformly charged $S_{D-1}$, i.e.
a hypersphere embedded in $D$ dimensions, with radius $R=L/2$ and with
\emph{constant} hypersurface charge density $\sigma_{D}=Q/\left(  \Omega
_{D}R^{D-1}\right)  $. \ An orthogonal projection of the charge, $dQ$, from a
hyper-cylindrical \textquotedblleft ribbon\textquotedblright\ of revolution
about a diameter of the hypersphere onto the underlying surrounded segment
$dz$ of the diameter, gives precisely $dQ/dz=\lambda_{D}\left(  z\right)  $
for $-L/2\leq z\leq L/2$. \ This follows directly from $dQ=\sigma_{D}dA$,
where the hypersurface \textquotedblleft area\textquotedblright\ $dA$ of the
hyper-ribbon that surrounds $dz$ is $dA=\Omega_{D-1}~R^{D-2}\sin^{D-2}%
\theta~ds$ with $ds=Rd\theta$, $z=R\cos\theta$, and $\sin\theta=\sqrt
{1-z^{2}/R^{2}}$. \ This construction is easily visualized for $D=2$ and $3$.

We plot the densities for various integer $D$, from $D=2$ up to $D=11$, shown
respectively as the lower to upper curves (for $z=0$) in the following graph.%
%TCIMACRO{\FRAME{dtbpFU}{4.1191in}{2.7408in}{0pt}{\Qcb{Figure 22: \ Linear
%charge density $\lambda_{D}\left(  z\right)  $ leading to ellipsoidal
%equipotentials, in various dimensions.}}{}{eurojphyslinesegmentsv2__24.pdf}%
%{\special{ language "Scientific Word";  type "GRAPHIC";
%maintain-aspect-ratio TRUE;  display "USEDEF";  valid_file "F";
%width 4.1191in;  height 2.7408in;  depth 0pt;  original-width 4.5857in;
%original-height 3.0415in;  cropleft "0";  croptop "1";  cropright "1";
%cropbottom "0";
%filename 'EuroJPhysLineSegmentsV2__24.pdf';file-properties "XNPEU";}} }%
%BeginExpansion
\begin{center}
\includegraphics[
height=2.7408in,
width=4.1191in
]%
{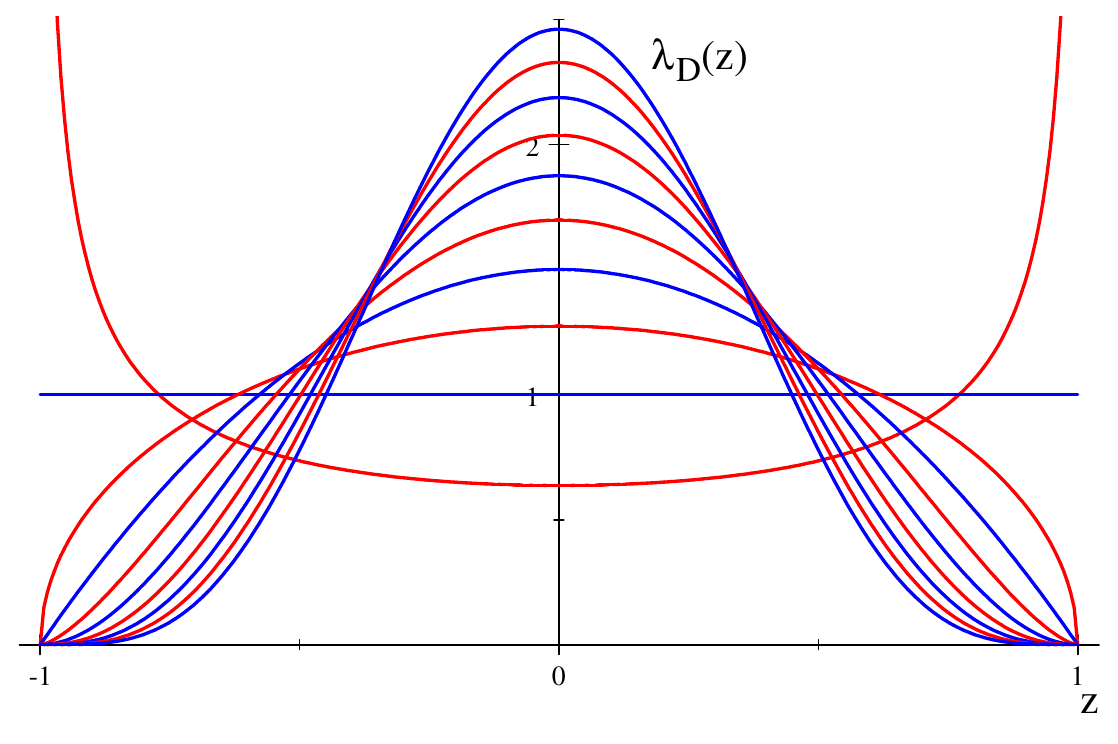}%
\\
Figure 22: \ Linear charge density $\lambda_{D}\left(  z\right)  $ leading to
ellipsoidal equipotentials, in various dimensions.
\end{center}
%EndExpansion

In this graph the 2D case is the only one which is intuitive in the sense that
the infinitesimal pieces of charge repel each other so we would expect excess
charge/length to be pushed towards the ends of a finite length, equipotential
segment of real conductor with small but nonzero transverse size, all parts of
which are at the same finite $\Phi$. \ For the idealized line segment in 3D
this is not so --- the charge density is uniform --- but then the segment
itself is at an infinite $\Phi$ so physical intuition based on real conductors
is perhaps difficult to apply in this situation \cite{Footnote5}. \ In higher
dimensions, the result is even \emph{more} counter-intuitive for idealized
line segments. \ The charge distribution is peaked at the center of the
segment. \ Of course, in making these statements, we are assuming the physical
properties of the idealized segment itself can be completely understood
mathematically by taking the limit of the surrounding equipotential ellipsoids
of revolution as their girth goes to zero.

To gain more insight about the behavior of the charge distribution and the
resulting potential, we may think of $D$ in (\ref{ChargeDistribution}) as a
continuous variable --- not just an integer --- to be used as a regulating
parameter by which the $D=3$ case can be approached as a limit. \ This point
of view shows that all $D<3$ behave intuitively, with $\lambda_{D}\left(
z\right)  $ peaked at the ends of the segment. \ The extreme case is $D=1$
where all charge is located \emph{only} at the ends of the segment.
\ Moreover, for all $D<3$ the potential along the segment, as given by
(\ref{3PointPotential}), is a \emph{finite} constant, so there is nothing
pathological about the potential that might obviate physical intuition based
on real conductors. \ 

On the other hand, all $D>3$ behave counter-intuitively, with $\lambda
_{D}\left(  z\right)  $ peaked at the center of the segment and vanishing at
the segment ends. \ And for all $D>3$ the potential along the segment, as
given by (\ref{3PointPotential}), is \emph{infinite}, so real-world physical
intuition is not guaranteed to be reliable for these idealized situations.
\ The uniformly charged 3D case, also with infinite $\Phi$ along the segment,
acts as a separatrix between intuitive and counter-intuitive charge
distributions. \ A plot of $\lambda_{D}\left(  z\right)  $ as a surface over
the $\left(  z,D\right)  $ plane helps to visualize these features.%
%TCIMACRO{\FRAME{dtbpFU}{6.7348in}{4.4775in}{0pt}{\Qcb{Figure 23: \ The
%$\lambda_{D}\left(  z\right)  $ surface as a continuous function of both $z$
%and $D$ for a straight-line segment from $z=-1$ to $z=1$. Constant density for
%a uniformly charged $D=3$ segment is represented by a thick red line.}}%
%{}{eurojphyslinesegmentsv2__25.pdf}{\special{ language "Scientific Word";
%type "GRAPHIC";  maintain-aspect-ratio TRUE;  display "USEDEF";
%valid_file "F";  width 6.7348in;  height 4.4775in;  depth 0pt;
%original-width 6.8351in;  original-height 4.5343in;  cropleft "0";
%croptop "1";  cropright "1";  cropbottom "0";
%filename 'EuroJPhysLineSegmentsV2__25.pdf';file-properties "XNPEU";}} }%
%BeginExpansion
\begin{center}
\includegraphics[
height=4.4775in,
width=6.7348in
]%
{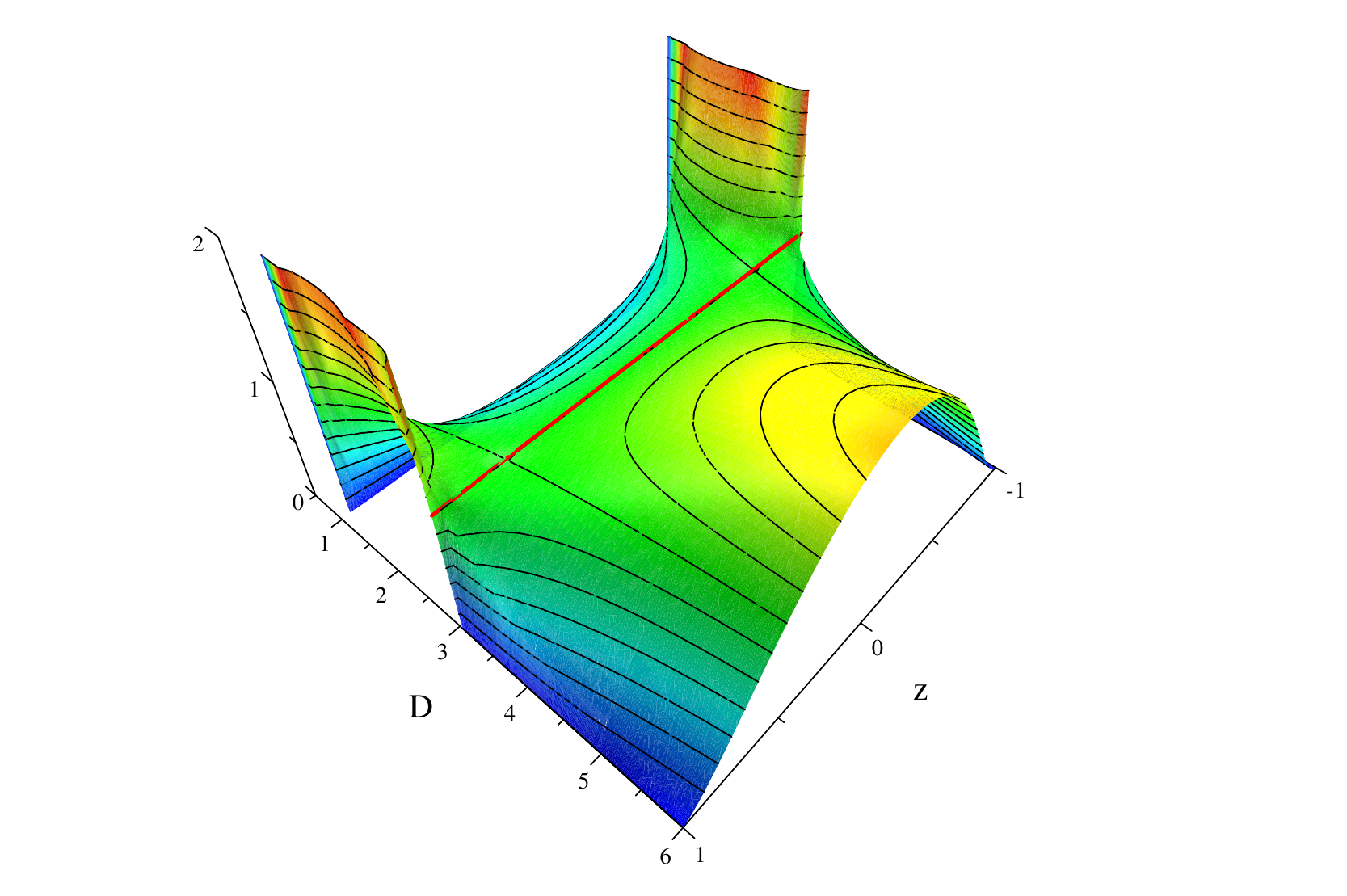}%
\\
Figure 23: \ The $\lambda_{D}\left(  z\right)  $ surface as a continuous
function of both $z$ and $D$ for a straight-line segment from $z=-1$ to $z=1$.
Constant density for a uniformly charged $D=3$ segment is represented by a
thick red line.
\end{center}
%EndExpansion

We close this Section with a discussion of the direct calculation of the
potential and the electric field as linear superpositions of the $d\Phi$'s and
$d\overrightarrow{E}$'s due to infinitesimal bits of charge along the segment,
$\lambda_{D}\left(  z\right)  dz$, for a line segment in $D$ dimensions. \ In
terms of the obvious rectilinear coordinates, as shown in Figure 14 of Section
4, the results are%
\begin{equation}
\Phi_{\text{line}}\left(  w,h\right)  =k\int_{-L/2}^{L/2}\lambda_{D}\left(
z\right)  \left(  \frac{1}{\sqrt{\left(  h-z\right)  ^{2}+w^{2}}}\right)
^{D-2}~dz\ ,
\end{equation}%
\begin{equation}
\overrightarrow{E}_{\text{line}}\left(  w,h\right)  =\left(  D-2\right)
k\int_{-L/2}^{L/2}\lambda_{D}\left(  z\right)  \left(  \frac{1}{\sqrt{\left(
h-z\right)  ^{2}+w^{2}}}\right)  ^{D-1}\widehat{n}~dz\ ,
\end{equation}
where $w$ is the transverse distance from the segment and $h$ is the
$z$-coordinate of the observation point, and where $\widehat{n}=\frac
{\overrightarrow{r}-z\widehat{z}}{\left\vert \overrightarrow{r}-z\widehat{z}%
\right\vert }$ is a unit vector pointing from the location $z\widehat{z}$\ of
the bit of charge on the segment to the observation point $\overrightarrow{r}%
$. \ Now, using the charge distribution (\ref{ChargeDistribution}) that
produces ellipsoidal equipotentials, we eventually obtain%
\begin{equation}
\Phi_{\text{line}}\left(  w,h\right)  =\frac{2k\lambda}{\Omega_{D}%
/\Omega_{D-1}}\left(  \frac{2/L}{\sqrt{\sin\theta_{\text{b}}\sin
\theta_{\text{t}}}}\right)  ^{D-3}\int_{\theta_{\text{b}}}^{\theta_{\text{t}}%
}\left(  \sqrt{\sin\left(  \vartheta-\theta_{\text{b}}\right)  \sin\left(
\theta_{\text{t}}-\vartheta\right)  }\right)  ^{D-3}\frac{1}{\sin\vartheta
}~d\vartheta\ , \label{PotentialThetaIntegral}%
\end{equation}%
\begin{equation}
\overrightarrow{E}_{\text{line}}\left(  w,h\right)  =\frac{2k\lambda}{w}%
\frac{\left(  D-2\right)  }{\Omega_{D}/\Omega_{D-1}}\left(  \frac{2/L}%
{\sqrt{\sin\theta_{\text{b}}\sin\theta_{\text{t}}}}\right)  ^{D-3}\int%
_{\theta_{\text{b}}}^{\theta_{\text{t}}}\left(  \sqrt{\sin\left(
\vartheta-\theta_{\text{b}}\right)  \sin\left(  \theta_{\text{t}}%
-\vartheta\right)  }\right)  ^{D-3}\widehat{n}~d\vartheta\ ,
\label{ElectricFieldThetaIntegral}%
\end{equation}
where we have changed integration variables from $z$ to an angle $\vartheta$
according to%
\begin{align}
z  &  =h-w\cot\vartheta\ ,\ \ \ dz=\frac{w}{\sin^{2}\vartheta}~d\vartheta\ ,\\
r  &  =\sqrt{\left(  h-z\right)  ^{2}+w^{2}}=\sqrt{w^{2}\cot^{2}%
\vartheta+w^{2}}=\left\vert \frac{w}{\sin\vartheta}\right\vert \ ,\\
\sqrt{\frac{1}{4}~L^{2}-z^{2}}  &  =\frac{L}{2}\sqrt{1-\frac{4}{L^{2}}\left(
h-w\cot\vartheta\right)  ^{2}}=\sqrt{\frac{w^{2}\sin\left(  \vartheta
-\theta_{\text{b}}\right)  \sin\left(  \theta_{\text{t}}-\vartheta\right)
}{\sin^{2}\vartheta\sin\theta_{\text{b}}\sin\theta_{\text{t}}}}\ .
\label{CrucialID}%
\end{align}
Note that $\vartheta$ here is the polar angle \emph{measured from the location
of the bit of charge}, and in general this is not the $\theta$ shown in Figure
14 of Section 4. \ Nonetheless, $\theta_{\text{b}}$ and $\theta_{\text{t}}$
\emph{are} the angles shown in Figure 14. \ 

These last relations, in particular (\ref{CrucialID}) and
(\ref{ElectricFieldThetaIntegral}), are useful to establish \emph{the
direction} \emph{of the electric field} \emph{without actually performing the
integration}. \ To verify this statement, let $\vartheta=\psi+\frac{1}%
{2}\left(  \theta_{\text{b}}+\theta_{\text{t}}\right)  $ in
(\ref{ElectricFieldThetaIntegral})\ to find%
\begin{align}
\overrightarrow{E}_{\text{line}}\left(  w,z\right)   &  =\frac{2k\lambda}%
{w}\frac{\left(  D-2\right)  }{\Omega_{D}/\Omega_{D-1}}\left(  \frac
{2/L}{\sqrt{\sin\theta_{\text{b}}\sin\theta_{\text{t}}}}\right)  ^{D-3}%
\int_{\frac{\theta_{\text{b}}-\theta_{\text{t}}}{2}}^{\frac{\theta_{\text{t}%
}-\theta_{\text{b}}}{2}}\left(  \sqrt{\sin\left(  \psi+\frac{\theta_{\text{t}%
}-\theta_{\text{b}}}{2}\right)  \sin\left(  \frac{\theta_{\text{t}}%
-\theta_{\text{b}}}{2}-\psi\right)  }\right)  ^{D-3}\widehat{n}\left(
\psi\right)  d\psi\ .\nonumber\\
&  \label{ELineInD}%
\end{align}
The integration here weights $\widehat{n}\left(  \psi\right)  $ by an
\emph{even} function of $\psi$, with the range of integration symmetric about
$\psi=0$, so it follows that the resulting direction of $\overrightarrow{E}%
_{\text{line}}\left(  w,h\right)  $ will be proportional to $\widehat{n}%
\left(  \psi=0\right)  =\frac{1}{\left\vert \widehat{r}_{\text{t}}%
+\widehat{r}_{\text{b}}\right\vert }\left(  \widehat{r}_{\text{t}}%
+\widehat{r}_{\text{b}}\right)  $. \ That is to say, at any observation point
around the segment the direction of $\overrightarrow{E}_{\text{line}}$ bisects
the angle formed by the two lines from the end points of the segment to the
observation point, thereby confirming that the equipotentials are ellipsoids
of revolution about the segment.

Note that this direct calculation of $\overrightarrow{E}_{\text{line}}$ using
the integral expression (\ref{ELineInD}) generalizes prior calculations\ for
$D=3$ where the charge density is uniform, as given in
\cite{KelvinTait,Routh,Durand53,Durand64,GHR}, to cases where the charge
distribution on the segment is \emph{not} uniform for $D\neq3$. \ We leave it
as an exercise for the reader to evaluate the integral in (\ref{ELineInD}) to
obtain explicit expressions, and to show that these are in agreement with
$\overrightarrow{E}_{\text{line}}=-\overrightarrow{\nabla}\Phi_{\text{line}}$.
\ Even before attempting to evaluate the integral in (\ref{ELineInD}),
however, it should be evident that a determination of $\overrightarrow{E}%
_{\text{line}}$ by taking the gradient of the potential is easier to carry through.

\section{Generalizations}

Having found the equipotentials for various charged line segments, we have in
hand a set of solutions for a variety of electrostatic boundary value
problems. \ If all the charge from the line segment is moved outward along
electric field lines and placed on a single surface selected from among the
equipotentials, in such a way as to preserve $\Phi$ on that surface, then the
surrounding equipotentials are unchanged as well. \ This is widely known, and
was in fact a feature emphasized by Green \cite{Green}. \ The charge
distribution on the selected surface, as required to carry out this feat, is
of course given by the electric field normal to that surface as provided by
the original charged line segment solution. \ In particular, ellipsoidal
equipotentials in any $D$ yield hypersurface charge densities that are
expressible as elementary functions. \ The results are straightforward
generalizations of the well-known 3D ellipsoidal case. \ 

In fact, it is not even necessary for the hyperellipsoids to be surfaces of
revolution. \ For a general ellipsoid, as defined in $D$ dimensions by
\begin{equation}
\sum_{n=1}^{D}\frac{x_{n}^{2}}{a_{n}^{2}}=1\ , \label{EllipsoidD}%
\end{equation}
the charge distribution on that hyperellipsoid such that it is an
equipotential is given by%
\begin{equation}
\sigma_{D}\left(  \overrightarrow{r}\right)  =\frac{Q}{\Omega_{D}\left(
\prod\limits_{n=1}^{D}a_{n}\right)  }\frac{1}{\sqrt{\sum_{n=1}^{D}x_{n}%
^{2}/a_{n}^{4}}}\ . \label{EllipsoidSurfaceCharge}%
\end{equation}
This can be established by a straightforward adaptation of the argument that
applies to the 3D case, as given for example in Smythe \cite{Smythe}, Chapter
5, \S 5.00-\S 5.02.

If this charged hyperellipsoidal surface is \textquotedblleft
squashed\textquotedblright\ to obtain an equipotential hyperdisk (see Smythe
\S 5.03 for the 3D case), say by letting $a_{D}\rightarrow0$ while maintaining
the constraint (\ref{EllipsoidD}), then the hypersurface charge density on the
resulting hyperdisk (counting charge on both sides of the disk) becomes%
\begin{equation}
\sigma_{D}\left(  r\right)  =\frac{2Q}{\Omega_{D}\left(  \prod\limits_{n=1}%
^{D-1}a_{n}\right)  }\frac{1}{\sqrt{1-\sum_{n=1}^{D-1}x_{n}^{2}/a_{n}^{2}}}\ .
\end{equation}
This charge distribution may now be projected along any of the remaining
unsquashed principal axes of the original ellipsoid by elementary
integrations. \ The final result is exactly the linear charge distribution
given by (\ref{ChargeDistribution}), namely,
\begin{equation}
\lambda_{D}\left(  x_{k}\right)  =\frac{Q}{a_{k}}~\frac{\Omega_{D-1}}%
{\Omega_{D}}~\left(  1-\frac{x_{k}^{2}}{a_{k}{}^{2}}\right)  ^{\frac{D-3}{2}%
}\ ,
\end{equation}
where the projection has been made onto the $k$th axis. \ Alternatively, the
surface charge distribution (\ref{EllipsoidSurfaceCharge}) of the unsquashed
ellipsoid may be projected directly without being squashed, again by
elementary integrations, to obtain the same result for $\lambda_{D}$.

\section{Summary}

In the context of a familiar subject --- electrostatics --- we have carried
out several elementary calculations in various numbers of spatial dimensions,
$D$, to encourage students to think more critically about the role played by
$D$. \ 

Although for $D=3$ a uniformly charged straight-line segment gives rise to an
electric potential $\Phi$ whose equipotential surfaces are prolate ellipsoids
of revolution about the segment, with the ends of the segment providing the
foci of the ellipsoid, this is a very special case. \ If $D\neq3$, uniformly
charged segments do not produce ellipsoidal equipotentials. \ If $D\neq3$, a
non-uniform distribution of charge is required to produce equipotentials that
are ellipsoidal about a straight-line segment of charge. \ 

We have illustrated these $D\neq3$ features in detail for $D=2$ and $D=4$, and
we have provided a framework as well as explicit formulas to carry through the
same level of detail for any $D$. \ We have shown how $D=2$ is the only
ellipsoidal equipotential case which is intuitive in the sense that the
associated linear charge distribution has maxima at the ends of the segment,
as one might naively expect. \ In contrast, we have shown that the
distribution of charge needed to produce ellipsoidal equipotentials is
counter-intuitive for $D>3$, becoming all the more so as $D$ is increased, in
the sense that the requisite charge distribution has an absolute maximum at
the center of the segment and vanishes at its ends. \ 

For $D=3$ and $D=4$, the potential surface plots were found to have similar
features that can be distinguished only by careful inspection. \ Differences
in $\Phi$ between uniformly charged line segments and those giving rise to
ellipsoidal equipotentials were shown to be somewhat subtle. \ Distinctions
between the two cases were more easily drawn by examining the corresponding
electric fields close to the segment, or, relatedly, by directly comparing the
charge densities.

We also took the opportunity to illustrate, albeit briefly, how continuous $D
$ can be viewed as a mathematical tool to regulate singular behavior and to
interpolate between intuitive and counter-intuitive charge distributions. \ In
the course of our discussion, we employed some basic geometrical ideas at an
elementary level, such as in the determination of the direction of the
electric field associated with ellipsoidal equipotentials.

Finally, we indicated how the line segment results may be generalized to solve
a class of electrostatic boundary value problems by distributing the charge on
hypersurfaces, rather than straight lines, while maintaining the line segment
equipotentials outside the charged hypersurface.\bigskip

\textbf{Acknowledgement:} \ We thank T S Van Kortryk for helpful discussions
of the topics presented here.


\begin{thebibliography}{99}
\bibitem {Footnote0}Throughout our discussion we use rectangular Cartesian,
spherical polar, and cylindrical coordinates, in view of student familiarity
with these coordinate systems, even though prolate ellipsoids of revolution
are more naturally described in terms of prolate spheroidal coordinates. \
\bibitem {Footnote1}It is tempting to introduce the variable $t=r_{+}-r_{-}$,
the coordinate locally orthogonal to $s$ in a prolate spheroidal coordinate
system. \ This would make the result for $\overrightarrow{E}$\ just as compact
as that for $\Phi_{\text{line}}$ when the equipotentials are ellipsoidal.
\ However, the results are sufficiently manageable in terms of rectangular
Cartesian coordinates, at least for graphical purposes, that there is no real
numerical advantage achieved by using prolate spheroidal coordinates. \
\bibitem {Footnote2}Beginning students can appreciate why the normal to a
prolate ellipsoidal surface bisects the angle formed by the two vectors from
the foci to any point on the ellipsoid by thinking about the law of reflection
applied to a light ray that strikes an ellipsoidal mirror while en route from
one focus to the other. \ Alternatively, a few banked shots on any good
ellipsoidal billiard table should clarify the situation.
\bibitem {Footnote3}Note that a careful limit of (\ref{PhiD}) is required to
obtain the correct potential for $D=2$. \ Just setting $D=2$ in (\ref{PhiD})
would give a constant $\Phi$.
\bibitem {Footnote4}The form in (\ref{Alternate3PointPotential}) is also more
useful to obtain the correct result for $D=2$ since the form in
(\ref{3PointPotential}) requires taking a limit $D\rightarrow2$ to obtain a
nontrivial result.
\end{thebibliography}
\end{document}